\documentclass{aa}

\usepackage{psfig,graphicx,natbib}

\newcommand{\pc}{\>{\rm pc}}

\newcommand{\ha}{H$\alpha$\,}
\newcommand{\oxy}{[\hbox{O\,{\sc iii}}]\,}

\begin{document}


\titlerunning{Gaseous outflows in edge-on Seyfert galaxies}
\authorrunning{Robitaille et al.}

\title{The morphology of minor axis gaseous outflows in edge-on Seyfert 
galaxies\thanks{Based on observations collected at the Centro Astron\'omico 
Hispano Alem\'{a}n (CAHA) at Calar Alto, operated jointly by the Max-Planck 
Institut f\"ur Astronomie and the Instituto de Astrof\'{\i}sica de 
Andaluc\'{\i}a (CSIC)}}

\author{T. P. Robitaille\thanks{also summer student and visitor at the Space 
                          Telescope Science Institute}
       \inst{1}
       \and
       J. Rossa
       \inst{2,3}
       \and
       D. J. Bomans
       \inst{4}
       \and
       R. P. van der Marel
       \inst{2}
       }
\institute{SUPA, School of Physics and Astronomy, University of St Andrews, 
           North Haugh, KY16 9SS, St Andrews, United Kingdom\\
           \email{tr9@st-andrews.ac.uk}
      \and
          Space Telescope Science Institute, 3700 San Martin Drive, 
          Baltimore, MD 21218, U.S.A.\\
          \email{marel@stsci.edu}
      \and
          Department of Astronomy, University of Florida, 211 Bryant Space 
          Science Center, P.O.Box 112055, Gainesville, FL 32611, U.S.A.\\
          \email{jrossa@astro.ufl.edu}
      \and
          Astronomisches Institut, Ruhr-Universit\"at Bochum, 
          Universit\"atsstrasse 150/NA7, D-44780 Bochum, Germany\\
          \email{bomans@astro.rub.de}
          }

\offprints{J.~Rossa}

\date{Received 19 April 2006 / Accepted 20 December 2006}


\abstract{Spiral galaxies often have extended outflows that permeate
beyond the region of the disk. Such outflows have been seen both in
starburst galaxies, actively star forming galaxies and galaxies with
an AGN. In the latter galaxies it is unknown whether the large-scale
outflows are driven by star formation activity or purely by the active
nucleus.}{The aim of our investigation is to study the frequency of
extended minor-axis outflows in edge-on Seyfert galaxies to
investigate the role of the AGN, the circumnuclear environment and
star formation activity within the disk regions, and their importance
for IGM enrichment on large scales.}{We obtained optical narrowband
imaging observations of a distance limited, northern hemisphere sample
of 14 edge-on Seyfert spiral galaxies. Because of the distance-limited
nature of the sample, it is restricted to relatively low-luminosity
Seyfert galaxies. The data were obtained with BUSCA attached to the
2.2m telescope at the Calar Alto observatory. Narrowband imaging in
two different ionizational stages (\ha and \oxy) was performed to
attempt a discrimination between processes associated with the active
nucleus and those connected to star forming activity within the
disk. The median 3-$\sigma$ sensitivities for detection of
high-latitude extended emission in the sample galaxies are
$3.6\times10^{-17}$\,ergs\,s$^{-1}$\,cm$^{-2}$\,arcsec$^{-2}$ for the
\ha images and
$6.9\times10^{-17}$\,ergs\,s$^{-1}$\,cm$^{-2}$\,arcsec$^{-2}$ for the
\oxy images. We use the data to study the distribution of extraplanar
emission with respect to the AGN and the underlying disk \hbox{H\,{\sc
ii}} regions.}{The \ha morphology of the Seyfert galaxies is usually
complex, but only in three out of 14 galaxies did we find evidence for
minor axis disk outflows. At the sensitivity of our observations \oxy
emission is generally detected only in the nuclear region. For Ark\,79
we present the first evidence of a secondary nuclear component, best
visible in the \oxy image, which has a linear separation from the
primary nucleus of about 850\,pc.}{Overall, our results show that
extraplanar emission of similar brightness and extent as in the
previously known cases of NGC\,3079 and NGC\,4388 is not common in
Seyfert galaxies of otherwise similar properties. Comparison with our
previous results shows that for nearby edge-on spiral galaxies star
formation may be a more powerful mechanism for producing DIG than AGN
activity. While in general AGN activity undoubtedly plays some role in
driving minor-axis outflows, this probably requires higher AGN
luminosities than are encountered in our small distance-limited
sample.}


    \maketitle


\section{Introduction}
\label{s:intro}

The investigation of minor axis gaseous outflows in external galaxies by 
narrowband imaging is well established for starburst galaxies 
\citep[e.g.,][]{leh95,leh96a} and also for normal ({\em non-starburst}) 
galaxies \citep[e.g.,][]{ran96,ros00,mil03,ros03b}. In starburst galaxies the 
presence of gaseous outflows is a common feature, where the gas and energy 
are being transported to the ambient intergalactic medium (IGM) from the 
nuclear region via starburst driven winds. In normal galaxies the 
presence of minor axis gaseous outflows is not as ubiquitous, but is however 
still a common phenomenon for galaxies exceeding a certain threshold of star 
formation activity in the underlying galaxy disk \citep{ran96,ros03a}. 

Previous studies on minor axis outflows in Seyfert galaxies have been
performed mostly by \ha imaging \citep[e.g.,][]{pog89,col96}.
However, only a few galaxies have been studied in \oxy in the past
\citep[e.g.,][]{pog88a,pog88b,pog89,vei99a,vei03}, and these were
mostly relatively face-on galaxies. Apart from the recent study of
NGC\,4388 with the SUBARU telescope \citep{yos02}, most observations
had limited sensitivity, and only {\em nucleated} minor axis outflows
in \ha were detected.  Furthermore, the extent of the regions around
the galaxies studied was limited by detector size (i.e. limited field
of view). By contrast, the recent deep narrowband imaging in the
Seyfert galaxy NGC\,4388 revealed extended emission from minor axis
outflows up to distances of $|z|\sim$ 35\,kpc. Previous detection
rates of large scale minor axis outflows in edge-on Seyfert galaxies
were found to be about 27\% \citep{col96}. More recently,
\citet{vei03} presented results of a Taurus tunable filter study of
10, mostly face-on, starburst/AGN galaxies. Their overall detection
rate was 60\%, but when only counting the six Seyfert galaxies in
their sample, all of those showed large scale outflows.

The origin of very extended emission line regions is not quite clear yet. 
A likely origin for the extended emission in the Virgo Cluster galaxy 
NGC\,4388 may be ram pressure stripping \citep{yos04}, as had been previously 
suggested for the peculiar gas morphology in another Virgo cluster spiral, 
NGC\,4522 \citep{ken99,ken04}. Puzzling excitation structures of an outflow 
were reported for the galaxy NGC\,1482, using tunable filters \citep{vei02}, 
although on somewhat smaller scales than for the NGC\,4388 case. For a 
general review on galactic winds we refer the reader to the recent paper 
by \citet{vei05}.

In a recent investigation using \ha imaging of gaseous outflows in
nearby edge-on galaxies it was shown that there are differences
between starburst and non-starburst galaxies \citep{ros03a}. This is
manifested by their loci in a so-called diagnostic diagram for diffuse
emission, based on the strength of star formation rate per unit area
as a function of the far infrared flux ratio of 60 to 100 microns
(which is a proxy of the warm dust temperature). Galaxies with
higher star formation rate and warmer dust are more likely to show
outflows. In the \citet{ros03a} study, Seyfert galaxies were basically
excluded. The present paper fills this gap by studying the
gaseous outflows in a sample of Seyfert galaxies. We observed this
sample both in \ha and \oxy emission. The excitation mechanisms at
work can potentially be distinguished by the fact that the \oxy and
\ha emission will have different responses to ionizing contributions
of a hard power-law and from stellar radiation. Different excitation
mechanisms might therefore be distinguishable using \ha to \oxy ratio
maps.

The method of narrowband imaging has the advantage that excitation mechanisms 
can be studied in a much less time consuming manner than by longslit 
spectroscopy. The whole galaxy can be investigated in one single pointing per 
emission line. By contrast, longslit spectroscopy can only measure line 
ratios at specific cuts perpendicular or parallel to the major axis, where 
emission is already known from previous imaging. However, in many cases the 
presence or distribution of extended minor axis gaseous outflows in Seyfert 
galaxies is not known at all. A multi-filter narrowband imaging study of a 
homogeneous sample of Seyfert galaxies is therefore a natural first step.
With this approach we would like to investigate the role of the AGN versus 
the contribution to the IGM enrichment by disk outflows from strong 
\hbox{H\,\sc{ii}} regions. We can also study environmental effects such as 
past mergers \citep{wal96} and ram pressure stripping \citep{cay94}, by 
considering how disturbed the extended emission line regions are. 

The paper is structured in the following way. In Section~\ref{s:sampleobs} 
we present the sample, describe the observations and present the data
reduction techniques. In Section~\ref{s:res} we present the results for each 
individual galaxy, including a basic description of the observed morphology 
as seen in the broadband (e.g., $R$-band) and narrowband (\ha and \oxy) 
images. In Section~\ref{s:discuss} we discuss the results in terms of 
excitation maps and we attempt to discriminate between minor axis nuclear 
and disk outflows. Finally, in Section~\ref{s:sum} we present our summary.

\section{Sample, Observations and Data Reduction}
\label{s:sampleobs}

\subsection{The Seyfert Galaxy Sample}
\label{ss:sample}

Our sample is a distance-limited sample of nearby edge-on Seyfert 
galaxies. We selected the galaxies according to the following criteria: all 
targets have radial velocities of $v_{\rm rad}\leq5300\,{\rm km\,s^{-1}}$, 
and inclinations of $i\geq80^{\circ}$. The first selection criterion assures 
that sufficiently high spatial resolution is reached in order to resolve 
individual emission features, such as filaments. The galaxies in our sample 
have distances of $10.6$\,Mpc $\leq D \leq 73.0$\,Mpc, assuming a Hubble 
constant of $H_0 = 75$\,km\,s$^{-1}$\,Mpc$^{-1}$. This translates to spatial 
scales that can be resolved in the range of 50\,pc to 350\,pc for our 
targets. The inclination criterion is essential for detections of minor axis 
outflows, since in less inclined targets extended emission cannot be 
distinguished unambiguously from disk emission. 

The resulting sample consists of 14 northern hemisphere edge-on Seyfert 
galaxies that were selected from the electronic version of the 10th edition 
of {\em A Catalogue of Quasars and Active Nuclei}, published by \citet{ver01}.
We have chosen only galaxies classified as being Seyfert (type 1 and 2) 
galaxies, thus rejecting \hbox{H\,\sc{ii}} galaxies and LINERs. All relevant 
galaxies were initially inspected visually using Digitized Sky Survey (DSS) 
images to verify their edge-on orientation prior to making it into our 
final list. Our selection criteria are not too different from the \ha imaging 
study \citet{col96}. However, we found that several of the galaxies previously 
studied by \citet{col96} were far from being archetypical edge-on galaxies. 
In addition we include some known Seyferts, based on the search in 
\citet{ver01} that are not constrained in the \citet{col96} study. The 
selected sample consisted of 30 galaxies, visible from both 
hemispheres. We only observed the 14 galaxies visible from the northern 
hemisphere. In Table~\ref{t:basics} we list the basic parameters of these 
galaxies, such as coordinates, sizes, distances and magnitudes. 
\begin{table*}
\begin{flushleft}
\caption{Basic properties of the galaxy sample}
\label{t:basics}
\centering
\begin{tabular}{lcccccccc}
\hline\hline
Galaxy & R.A. (J2000.0) & Dec. (J2000.0) & cz & Distance & a$\times$b &
$m_B$ & $L_{\rm{FIR}}$ \\
& [hh mm ss.s] & [$\degr\,~\arcmin\,~\arcsec$] & [$\rm{km\,s^{-1}}$] & [Mpc]
& [$\arcmin\times\arcmin$] & [mag] & [$\rm{ergs\,s^{-1}}$] \\
(1) & (2) & (3) & (4) & (5) & (6) & (7) & (8) \\
\hline
Mrk\,993 & 01 25 31.4 & +32 08 11 & 4658 & 62.1 & 2.2$\times$0.7 & 13.96 &
2.78$\times10^{43}$\\
Mrk\,577 & 01 49 29.9 & +12 30 32 & 5221$^{a}$ & 69.6 & 1.3$\times$0.5 &
14.10 & ... \\
UGC\,1479 & 02 00 19.1 & +24 28 25 & 4927$^{a}$ & 65.7 & 1.3$\times$0.4 &
14.90 & 1.00$\times10^{44}$\\
Ark\,79 & 02 17 23.0 & +38 24 50 & 5254$^{a}$ & 70.1 & 0.8$\times$0.2 &
13.84 & ... \\
Mrk\,1040 & 02 28 14.6 & +31 18 41 & 4993 & 66.6 & 3.9$\times$0.8 & 14.74 &
1.66$\times10^{44}$\\
UGC\,2936 & 04 02 48.3 & +01 57 57 & 3813 & 50.8 & 2.5$\times$0.7 & 15.00 &
1.98$\times10^{44}$\\
NGC\,3079 & 10 01 57.8 & +55 40 47 & 1125 & 15.0 & 7.9$\times$1.4 & 11.43 &
1.39$\times10^{44}$\\
NGC\,3735 & 11 35 57.3 & +70 32 09 & 2692 & 35.9 & 4.2$\times$0.8 & 12.57 &
1.46$\times10^{44}$\\
NGC\,4235 & 12 17 09.9 & +07 11 29 & 2410 & 32.1 & 4.2$\times$0.9 & 12.64 &
4.67$\times10^{42}$ \\
NGC\,4388 & 12 25 46.7 & +12 39 44 & 2517 & 33.6 & 4.8$\times$0.9 & 11.91 &
1.61$\times10^{44}$\\
NGC\,4565 & 12 36 20.8 & +25 59 16 & 1282 & 17.1 & 15.9$\times$1.9 & 10.18 &
3.71$\times10^{43}$\\
NGC\,5866 & 15 06 29.6 & +55 45 48 & 769$^{a}$ & 10.3 & 4.7$\times$1.9 &
11.26 & 9.97$\times10^{42}$\\
IC\,1368 & 21 14 13.0 & +02 10 41 & 3912 & 52.2 & 1.1$\times$0.4 & 14.30 &
1.42$\times10^{44}$\\
UGC\,12282 & 22 58 55.5 & +40 55 53 & 5094 & 67.9 & 1.9$\times$0.5 & 14.70 &
8.06$\times10^{43}$\\
\hline
\end{tabular}
\end{flushleft}
Notes: Column~(1) lists the galaxy name. Columns~(2) and (3) list the
coordinates taken from the NASA Extragalactic Database (NED). Column~(4)
gives the radial velocity, measured from \hbox{H\,{\sc i}} data and taken
from NED. Data labeled with ``$a$'' refer to velocity data from optical
measurements. Column~(5) gives the corresponding distances assuming
$H_{0}=75\,$km\,s$^{-1}$\,Mpc$^{-1}$. Column~(6) lists the optical extent of
the major and minor axis taken from NED. Column~(7) lists the $B$-band
magnitude, also taken from NED. Finally, in column~(8) we list the FIR
luminosity, calculated from $F_{60}$ and $F_{100}$ (taken from NED) using
the equation in \cite{ros00}. For Ark\,79 and Mrk\,577 there are no IRAS
fluxes available, therefore $L_{\rm FIR}$ was not computed.
\end{table*}

The advantage of using a distance-limited sample is that the galaxies
can be studied with high angular resolution and sensitivity. It has
the disadvantage that only a relatively small volume is probed. Since
the most luminous AGN are rare, our sample is necessarily restricted
to ``low-luminosity'' Seyfert galaxies. Our study can therefore only
make statements about extended emission-line gas and minor-axis
outflows in Seyfert galaxies in this luminosity regime. The results
cannot necessarily be generalized to higher-luminosity Seyfert
galaxies, since it is likely that more luminous AGN are able to drive
stronger outflows. Our study is therefore complementary to possible
studies of more distant high-luminosity Seyferts. Such studies would
have different goals and limitations. For example, there would be less
spatial resolution and thus less sensitivity to detection of narrow
features (e.g., filaments). Moreover, the highest-luminosity AGN often
have ongoing star-formation, or are involved in a galaxy-galaxy
interaction. This makes it more difficult to establish whether any
outflow is driven by the AGN or by the starburst.

\subsection{Observations}
\label{ss:obs}

The sample was imaged in two observing runs. Optical $B$- and $R$-band 
images, and narrowband \ha and \oxy images were taken in the periods from 
October 28th to 30th 2002 and from April 5th to 6th 2003. We used the Bonn 
University Simultaneous Camera (BUSCA), attached to the 2.2m telescope at 
the Calar Alto Observatory. BUSCA \citep{rei99} is a four-channel CCD camera 
which allows simultaneous imaging of a $12\arcmin\times12\arcmin$ field of 
view in four different wavebands, each of which are separated by dichroic 
filters. The wavelength range covered by these four CCD cameras is $\lambda 
\leq$ 4300{\AA} (UV), 4300 to 5400{\AA} (Blue), 5400 to 7300{\AA} (Red), 
and $\lambda \geq$ 7300{\AA} (NIR). Three of the four CCDs are frontsize 
devices (CCD485 Lockheed Martin), with a pixel array of 4096$\times$4096 
pixels, and with a pixel size $15\mu$m $\times$ 15$\mu$m. The CCD for the UV 
channel is a backside thinned CCD485. The pixel scale in the $2\times2$ 
binning configuration used was 0\farcs35. Further details on BUSCA can be 
found at the BUSCA webpage\footnote{http://www.caha.es/guijarro/BUSCA/busca.html}.

\begin{figure*}
\centering
\includegraphics[angle=270,scale=0.7]{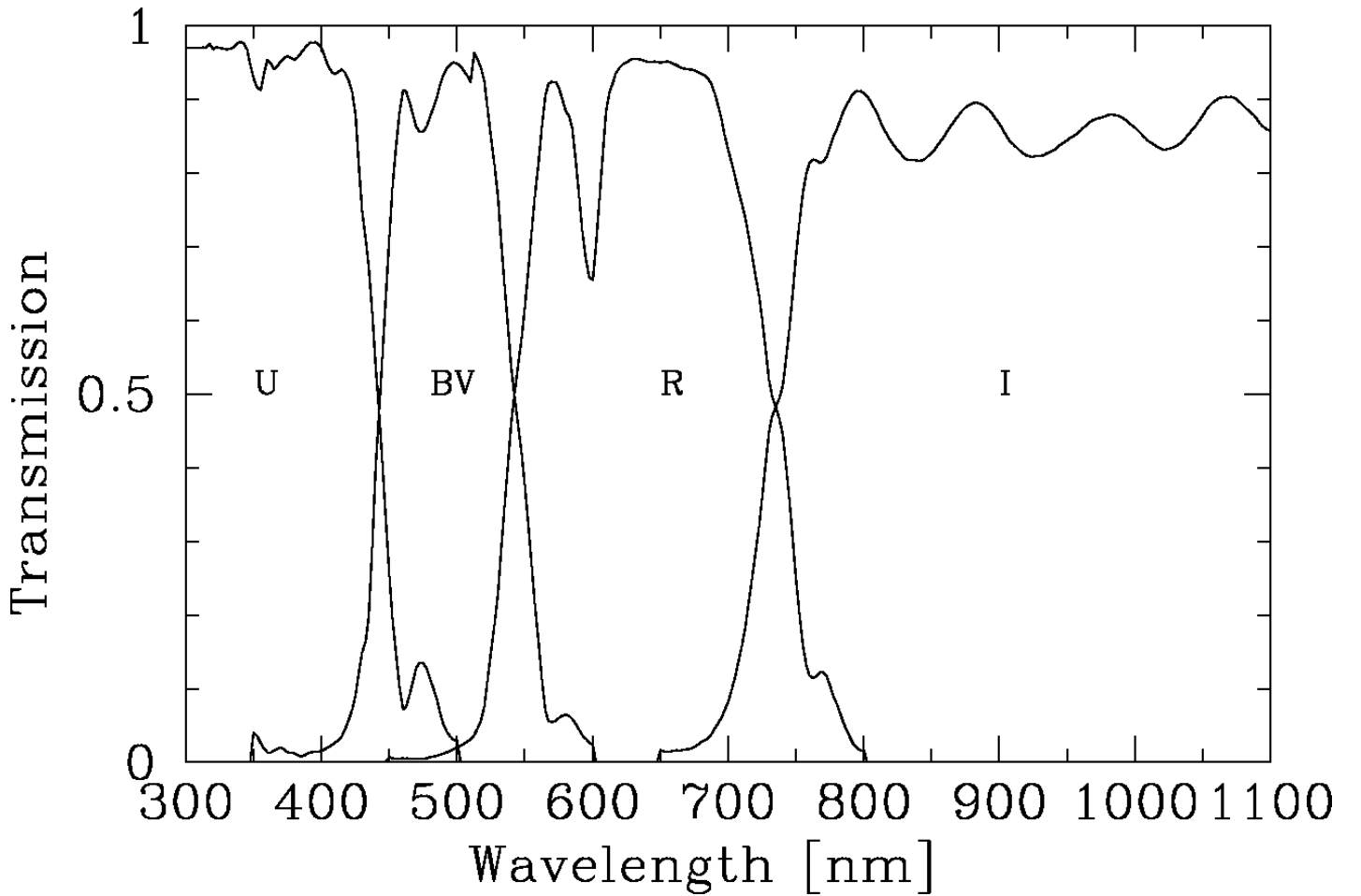}
\caption{Transmission curves for the four dichroic filters of the BUSCA
instrument. The positions in wavelength of standard broadband filters
are indicated. \label{f:dicroics}}
\end{figure*}

We used the H$\alpha$ and \oxy filters for channel 2 and 3, whereas
channel 1 and 4 were equipped with a UV and $I$-band filter.
The broadband images used for our analysis were
instead obtained in channel 2 and 3, subsequently, since their
bandpasses are located in the same wavelength region as the narrowband
filters. To avoid possible excessive straylight (Oliver Cordes for the
BUSCA team, priv. comm.), channels 2 and 3 were not equipped with
actual $B$ and $R$ broadband filters. Instead, the dichroic filters
already on the instrument adequately mimic the transmission of the
standard $B$ and $R$-bands. The transmission curves of these dichroic
filters are shown in Figure~\ref{f:dicroics}. 
Throughout this paper we
speak of $B$- and $R$-band images, when we refer to images obtained
with these dichroic filters.  Details on the narrowband filter
specifications such as central wavelength and FWHM are listed in
Table~\ref{t:obs}, which also presents a log of the observations that
includes observation dates, exposure times, and seeing values. Due to
the FWHM of the \ha filters, emission from the adjacent
[\hbox{N\,\sc{ii}}] doublet is also included in the bandpass. For the
remainder of the paper we always refer for simplicity to images
obtained with these filters as \ha images.
\onltab{2}{
\begin{table*}
\begin{flushleft}
\caption{Journal of observations}
\label{t:obs}
\centering
\begin{tabular}{lcccccccc}
\hline\hline
Galaxy & Date & $\lambda$ (\ha) & $\Delta\lambda$ (\ha) & $t_{\rm{exp}}$
(\ha \& R) & $\lambda$ ({\oxy}) & $\Delta\lambda$ ({\oxy}) & $t_{\rm{exp}}$
(\oxy \& B) & Seeing\\
 & [mm-dd-yyyy] & [\AA] & [\AA] & [sec] & [\AA] & [\AA] & [sec] & [arcsec] \\
(1) & (2) & (3) & (4) & (5) & (6) & (7) & (8) & (9)\\
\hline
Mrk\,993 & 10/28/2002 & 6667 & 76 & $6\times1200$ & 5066 & 104 &
$6\times1200$ \\
& & ... & ... & $3\times300+1\times600$ & ... & ... &
$3\times300+1\times600$ & 1.1 \\
Mrk\,577 & 10/29/2002 & 6667 & 76 & $6\times1200$ & 5066 & 104 &
$6\times1200$ \\
& & ... & ... & $3\times600$ & ... & ... & $3\times600$ & 0.9 \\
UGC\,1479 & 10/30/2002 & 6667 & 76 & $3\times2400$ & 5066 & 104 &
$3\times2400$ \\
& & ... & ... & $3\times600$ & ... & ... & $3\times600$ & 1.0 \\
Ark\,79 & 10/28,30/2002 & 6667 & 76 & $6\times1200$ & 5066 & 104 &
$6\times1200$ \\
& & ... & ... & $2\times600$ & ... & ... & $2\times600$ & 1.3\\
Mrk\,1040 & 10/30/2002 & 6667 & 76 & $1\times1200 + 2\times2400$ & 5066 &
104 & $1\times1200 + 2\times2400$ \\
& & ... & ... & $2\times600$ & ... & ... & $2\times600$ & 1.3\\
UGC\,2936 & 10/29/2002 & 6667 & 76 & $6\times1200$ & 5066 & 104 &
$6\times1200$ \\
& & ... & ... & $2\times600$ & ... & ... & $2\times600$ & 0.9\\
NGC\,3079 & 04/05/2003 & 6575 & 93 & $4\times1800$ & 5007 & 87 &
$4\times1800$ \\
& & ... & ... & $2\times600$ & ... & ... & $2\times600$ & 1.6 \\
NGC\,3735 & 04/06/2003 & 6575 & 93 & $3\times2400$ & 5007 & 87 &
$3\times2400$ \\
& & ... & ... & $2\times600$ & ... & ... & $2\times600$ & 1.3\\
NGC\,4235 & 04/05,06/2003 & 6575 & 93 & $2\times1200 + 2\times2400$ &
5007 & 87 & $2\times1200 + 2\times2400$  \\
& & ... & ... & $3\times600$ & ... & ... & $3\times600$ & 1.5\\
NGC\,4388 & 04/05/2003 & 6575 & 93 & $1\times2400$ & 5007 & 87 &
$1\times2400$ \\
& & ... & ... & $1\times600$ & ... & ... & $1\times600$ & 1.4\\
NGC\,4565 & 04/05/2003 & 6575 & 93 & $4\times1800$ & 5007 & 87 &
$4\times1800$\\
& & ... & ... & $2\times600$ & ... & ... & $2\times600$ & 1.6\\
NGC\,5866 & 04/05,06/2003 & 6575 & 93 &  $1\times1800 + 2\times2400$ &
5007 & 87 &  $1\times1800 + 2\times2400$  \\
& & ... & ... & $2\times600$ & ... & ... & $2\times600$ & 1.3\\
IC\,1368 & 10/28,29/2002 & 6667 & 76 & $6\times1200$ & 5066 & 104 &
$6\times1200$ \\
& & ... & ... & $3\times600$ & ... & ... & $3\times600$ & 1.1\\
UGC\,12282 & 10/30/2002 & 6667 & 76 & $3\times2400$ & 5066 & 104 &
$3\times2400$ \\
& & ... & ... & $2\times600$ & ... & ... & $2\times600$ & 0.9 \\
\hline
\end{tabular}
\end{flushleft}
Notes: Column~(1) lists the galaxy name. Column~(2) lists the date
of observation. Columns~(3) and (4) give the central wavelength of the \ha
filter and the FWHM, respectively. Column (5) gives the total \ha
exposure time on the top line for each object and the total $R$-band exposure
time on the second line. Columns~(6) and (7) list the central wavelength of
the \oxy filter and the FWHM, respectively. Note, that for the $B$- and $R$-band
we do not list a central wavelength and FWHM, as the dichroic filters were used
(see Section~\ref{ss:obs} and Figure~\ref{f:dicroics}). Column (8) gives the
total \oxy exposure time on the top line for each object and the total $B$-band
exposure time on the second line. Column~(9) lists the average seeing for each
galaxy. This was found by measuring the PSF FWHM on the final broadband and
narrowband images, and computing the mean of the four values obtained.
\end{table*}
\clearpage
}

\subsection{Data analysis}
\label{ss:reduc}
\subsubsection{Data reduction}
\label{sss:datared}

The data reduction was performed in the usual way using the 
IRAF\footnote{IRAF is distributed by the National Optical Astronomy 
Observatory, which is operated by the Association of Universities for 
Research in Astronomy, Inc. (AURA) under cooperative agreement with the 
National Science Foundation.} {\em ccdproc} package. A pedestal count level 
was determined and subtracted from each image using the overscan region. Ten 
to twenty bias frames were combined using the {\em zerocombine} task to 
create one master bias for each night and each of the four channels. 
The appropriate bias frames were subtracted from the flat-field frames and 
images to remove any two-dimensional structure remaining after the 
pedestal level subtraction. However, this bias subtraction was only performed 
on the \ha and $R$-band images. There were variability problems in the 
bias images from the CCD used for the \oxy and $B$-band images. Since the 
field-dependent variations of the bias is small, we decided not to use 
these images, and stick with a single constant pedestal instead. The 
flatfields were combined in groups of three to eight frames to give one 
normalized master flatfield per night, channel, and filter. These were used 
to flatfield the data. 

The sky background in the images was determined by manually selecting
in each frame three regions which showed no contamination from the
galaxy or from any stars, and measuring the median sky value. The
three values were then averaged to give a final sky background
value. If the scatter in the three values was above a threshold of
five counts, the regions used to calculate the medians were
changed. The sky background thus determined was subtracted. We
estimate the uncertainties in the background values to be less than a
few percent. In the next step the images were aligned by performing a
cross-correlation using the {\em xregister} task. First, a Fourier
cross-correlation was used to correct for large shifts with a nearest
pixel interpolation, then a finer discrete cross-correlation was used,
applying a polynomial interpolation. The image frames were then
combined using the {\em imcombine} task. Since at least three
narrowband frames were available for most of our galaxies, a median
combination was sufficient to eliminate the cosmic rays from the
narrowband observations.  For the broadband observations, only two
frames were acquired for most galaxies. These were combined using an
average combination along with the {\em crreject} rejection criterion
to eliminate the majority of the cosmic ray events. In one specific
case (NGC\,4388) only one frame was obtained in all filters, and the
{\em cosmicrays} package was used in order to eliminate the cosmic
rays.

Once cosmic-ray free images were obtained, the images from the different 
filterbands for a specific galaxy were aligned. This is a requirement for 
the continuum subtraction, which involves a scaling of different filter 
images. The alignment involved a small rotation, as the different channels 
did not have the exact same orientation (the measured difference was 
typically less than $1^\circ$). A polynomial interpolation was used to 
accommodate the difference in rotation angle. The images were then aligned 
using a discrete-cross correlation, once again applying a polynomial 
interpolation.

\subsubsection{Remaining data artifacts}
\label{ss:artifacts}

\begin{figure*}
\centering
\includegraphics[scale=0.8]{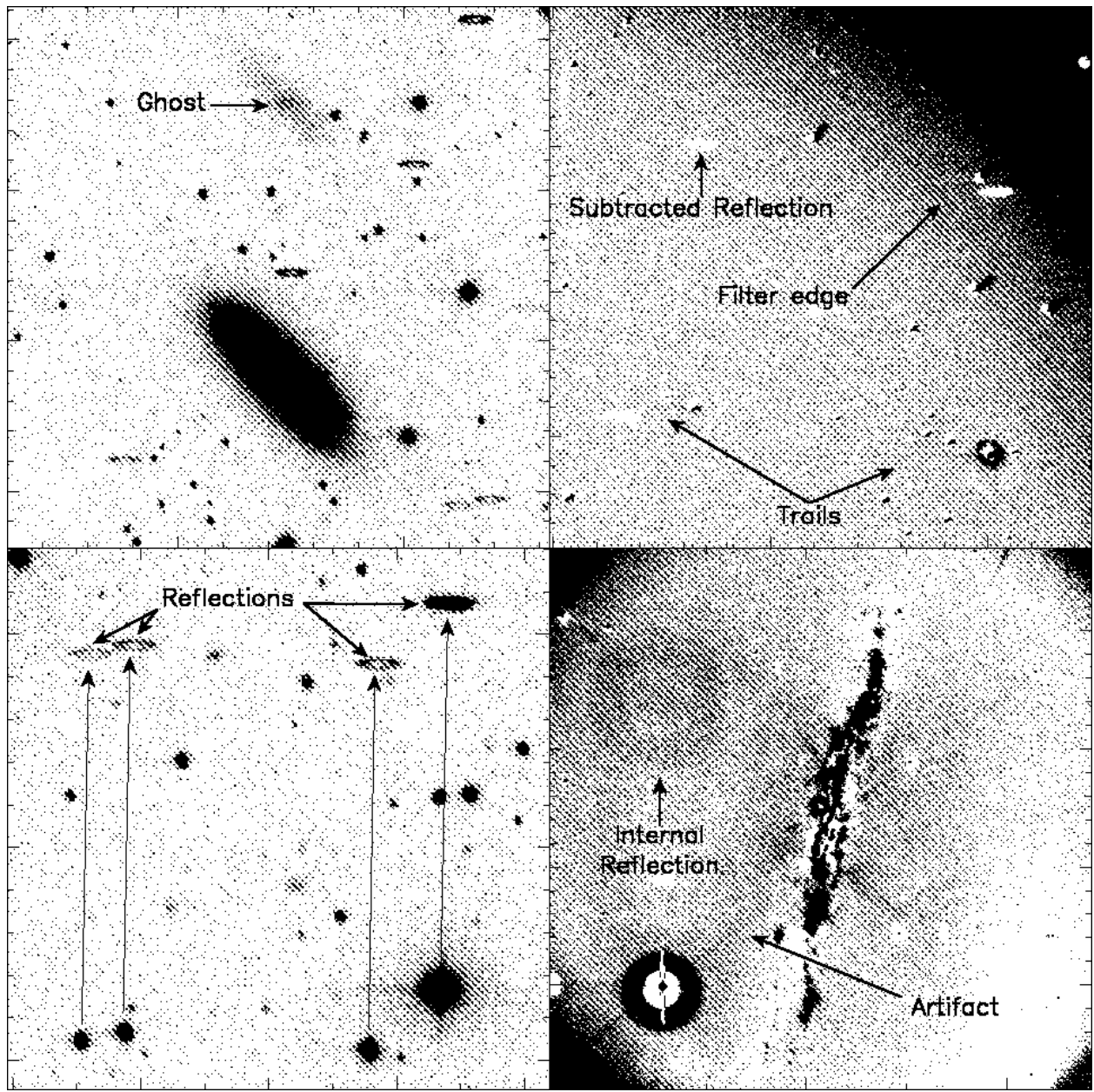}
\caption{Gallery of the various remaining image artifacts found in the BUSCA
data. A detailed description of these artifacts is given in
Section~\ref{ss:artifacts}. The images show some of the ``worst case''
examples. With proper care in the analysis our study was not generally
affected by any of these artifacts. \label{f:fourart}}
\end{figure*}

Besides the well known detector artifacts (e.g., those produced by cosmic-ray 
events), which can be corrected for by combining individual images, there 
are also instrumental artifacts caused by the optical system. This may leave 
additional artifacts (e.g., reflections) that we discuss in the following. 
Due to the optical outline of the BUSCA instrument (involving dichroic 
beamsplitters) in conjunction with the optical assembly of the telescope, 
there are several artifacts present in our images. We address these 
artifacts here and show examples in Figure~\ref{f:fourart}.

In all broadband images there are internal reflections from bright and 
moderately bright point sources (i.e. stars), which appear as slightly 
flattened, oval disks, with a central cavity (see upper left panel of 
Figure~\ref{f:fourart}). These reflections always have a constant distance 
from the source and appear $\approx 90\arcsec$ to the N of the stars in the 
$B$-band, and also appear $\approx 90\arcsec$ to the W and to the N in the 
$R$-band. The reflections are brightest in the $B$-band images. Since these 
reflections have a constant distance from their source (see lower left panel 
of Figure~\ref{f:fourart}), they are easy to identify and thus cannot be 
mistaken for any of the diffuse structures we are interested in. Furthermore, 
in the continuum subtracted images they appear as white oval disks (see upper 
right panel of Figure~\ref{f:fourart}) in contrast to the true galaxy 
features, which are shown in black in our ``inverted'' grey-scale images 
(see lower right panel of Figure~\ref{f:fourart}). Only in one case 
(UGC\,1479), was such an artifact directly superposed onto the southern part 
of the galaxy disk. Bright extended sources such as the galaxy bulge of 
course also produce ghost images at the same distance and orientation (N of 
the disk) in the $B$-band images (see upper left panel of 
Figure~\ref{f:fourart}). But because of the $90\arcsec$ offset, our analysis 
at the galaxies themselves is not affected by this. Other internal 
reflections around unusually bright stars in the field are also visible (see 
lower right panel of Figure~\ref{f:fourart}), reminiscent of the bright 
reflections seen in some of the Digitized Sky Survey images which have 
bright stars in the field. We have obviously taken extreme care not to 
mistake any of the associated diffraction spikes with emission filaments. 
Another artifact, affecting all CCDs to some extent, are the trails seen 
emerging from bright stars (see upper right panel of Figure~\ref{f:fourart}). 
These are so called charge transfer inefficiencies. 

It should also be noted that the narrowband filters obstructed the 
extreme outer boundaries of the field of view (see upper right 
panel of Figure~\ref{f:fourart}). This has no effects on any of our science 
cases, as most galaxies are much smaller than the field of view. 

\subsubsection{Narrowband emission maps}
\label{ss:narrow}

In order to generate continuum-free \ha and \oxy images, the broadband images 
were appropriately scaled and subtracted from the narrowband images. The 
$R$-band images were scaled and subtracted from the \ha images, and the 
$B$-band images were scaled and subtracted from the \oxy images. The scale 
factors were determined empirically.

\begin{figure*}
\centering
\includegraphics[angle=270,clip=t]{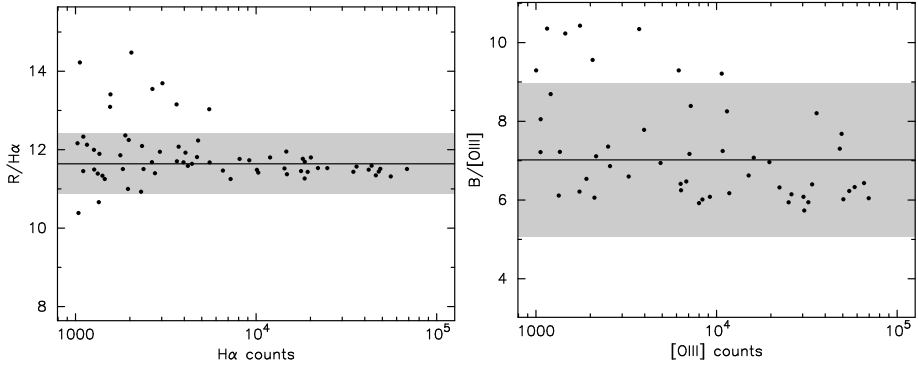}
\caption{Broadband to narrowband ratios of the foreground stars in the field
of view of UGC\,12282. The shaded region marks the $1\,\sigma$ scatter.
The R/\ha ratio is shown in the left panel, whereas the B/\oxy ratio is
shown in the right panel. The ratios are in counts with exposure times
scaled to correspond to the Ark\,79 values.
\label{f:scaleugc12282}}
\end{figure*}

In order to determine the $R$-band to \ha and $B$-band to \oxy ratios, we 
first performed aperture photometry on stars in both frames using the 
{\em apphot} package, attempting to include as many true stellar point 
sources as possible, while excluding background galaxies. 
Figure~\ref{f:scaleugc12282}
 shows the values of the ratio obtained for all 
the stars in the UGC\,12282 field. This field has the largest number of 
foreground stars, and therefore the ratios were best determined for this 
galaxy. Since some of the objects included in the photometry may still be 
non-stellar sources (e.g., faint background galaxies), we used the median 
ratio rather than the mean as an indicator of the overall ratio. The scatter 
at low count values is likely to be due to Poisson noise or in some cases 
non-stellar sources, but overall the $R$ to \ha ratio does seem well 
constrained. Unfortunately, the $R$ to \ha ratio was not as well constrained 
for most of the other galaxies, due to contamination from background galaxies 
and the lack of sufficient numbers of suitable stars in many of the frames. 
The $B$ to \oxy ratio was not tightly constrained in any of the galaxies.
Either way, the average color of foreground stars is probably not 
representative of the galaxies that we are interested in. This might bias 
the determination of the scale factors. So we decided to use a different 
approach instead, and to use aperture photometry results only as a 
consistency check.

We ultimately defined the broadband to narrowband ratio to be the ratio that 
minimizes the number of pixels with narrowband emission, whilst also 
minimizing the number of oversubtracted pixels. In practice this was done by 
selecting a patch of background sky and finding the standard deviation of 
the pixel counts within this region. We then considered all the pixels within 
a circular region, centered on each galaxy, and varied the broadband to 
narrowband ratio until the number of pixels with count values within one 
standard deviation from the mean background value was maximized. In order 
to verify this method we compared the ratio obtained to the method described 
above. We found the $R$ to \ha ratio for UGC\,12282 to be $11.23$ using the 
new algorithm, compared to $11.64\pm0.77$ for the initial method, which is in 
reasonable agreement. The $B$ to \oxy ratio was found to be $6.94$ using the 
new algorithm, compared to $7.02\pm1.94$ for the initial method, once again 
in good agreement. 

\begin{figure*}
\centering
\includegraphics[angle=270,clip=t]{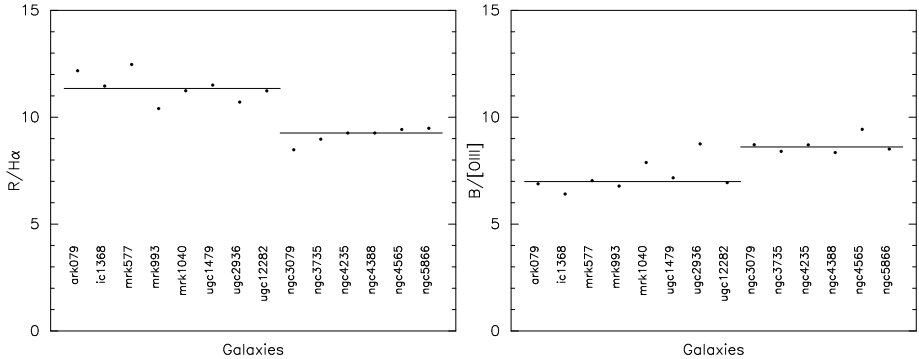}
\caption{Broadband to narrowband ratios for all galaxies in \ha (left)
and \oxy (right). The thin lines shows the median of the ratio for the
non-NGC galaxies, observed in the first run, and the median of the ratio for
the NGC galaxies, observed in the second run with a different filter. The
ratios are in counts with exposure times scaled to correspond to the Ark\,79
values. \label{f:scale}}
\end{figure*}

The final ratios used for each galaxy to construct the \ha and \oxy maps 
are shown in Figure~\ref{f:scale}, and are normalized to the individual 
exposure times. In the case of the $R$/\ha ratio, the values for the NGC 
galaxies are smaller than for the other galaxies. This agrees with the 
fact that the overall transmission of the filter used for the NGC galaxies 
is approximately $10$\% larger than the transmission of the filter used for 
the more remote galaxies. Reversely, the larger $B$/\oxy ratio for the NGC 
galaxies compared to the other galaxies is due to the fact that the \oxy 
filter used for the NGC galaxies has a transmission approximately $30$\% 
less than for the other galaxies. The relatively small variations in the 
ratio from galaxy to galaxy (observed with the same filter) confirm the 
accuracy of the method. Some of the remaining scatter may be due to the fact 
that the observing conditions were not photometric.

\subsubsection{Flux calibration}
\label{ss:flux}

The observations were partially carried out under non-photometric 
conditions. Therefore, no standards were acquired in the broadband 
channels. A flux calibration was performed using a narrowband image 
containing three calibration stars. The stars used were PG0231+051, 
PG0231+051A and PG0231+051D, and the standard magnitudes were taken from 
\citet{lan92}. The predicted flux of the stars in \ha and \oxy was found by 
normalizing a stellar spectrum to each of the known $B$-, $R$-, and $I$-band 
magnitudes in turn, each time extrapolating the flux to the \ha and \oxy 
wavelengths. The results using the three different broadband magnitudes were 
consistent with each other, and the calibration factors obtained for each of 
the three stars were also consistent. After taking the filter transmission 
into account, we found the calibration factors to be 
$(1.06\pm0.06)\times10^{-15}\,\rm{erg\,cm^{-2}\,count^{-1}}$ for 
the \ha images and $(1.01\pm0.06)\times10^{-15}\,\rm{erg\,cm^{-2}\,
count^{-1}}$ for the \oxy images. The errors represent the standard deviation 
of the values for the three stars and do not account for potential 
systematics.

The \ha and \oxy emission we observe from the various galaxies only 
spans a narrow range of wavelengths compared to the width of the 
filters. On the other hand, the flux from the calibration star was 
continuum emission, and therefore filled the whole filter. Therefore, 
before applying the calibration factor to the emission maps, we 
corrected for the filter transmission at the expected wavelength of the 
emission lines of the various galaxies. The expected wavelength of the \ha 
and \oxy lines were found by using the recession velocities of the galaxies. 
The \oxy emission line is actually a doublet, which means that the correction 
factor was found by taking the two lines into account (we assumed the 
$\lambda$4959/\oxy\,$\lambda$5007 ratio to be $1/3$).

The total \ha and \oxy flux of each galaxy was determined by summing
up the positive counts in a rectangular aperture enclosing the
galaxy. We subtracted any remaining contributions from stars enclosed
in the aperture. The \ha and \oxy fluxes for our sample are presented
in Table~\ref{t:fluxes}. 
\begin{table*}
\begin{flushleft}
\caption{\ha and \oxy fluxes and luminosities of the galaxies}
\label{t:fluxes}
\centering
\begin{tabular}{lcccc}
\hline\hline
Galaxy & $F$(H$\alpha$) & $L$(H$\alpha$) & $F$([O\sc{iii}]) &
$L$([O\sc{iii}]) \\
& [$\rm{ergs\,s^{-1}\,cm^{-2}}$] &  [$\log{L}$] & [$\rm{ergs\,s^{-1}\,
cm^{-2}}$] &  [$\log{L}$] \\
(1) & (2) & (3) & (4) & (5) \\
\hline
Mrk\,993 & 1.04$\times10^{-13}$ & 40.68 & 4.84$\times10^{-14}$ & 40.35\\
Mrk\,577 & 4.19$\times10^{-14}$ & 40.39 & 3.70$\times10^{-14}$ & 40.33\\
UGC\,1479 & 8.23$\times10^{-14}$ & 40.63 & 3.47$\times10^{-14}$ & 40.26\\
Ark\,79 & 1.67$\times10^{-13}$ & 40.99 & 1.22$\times10^{-13}$ & 40.86\\
Mrk\,1040 & 6.82$\times10^{-13}$ & 41.56 & 1.97$\times10^{-13}$ & 41.02\\
UGC\,2936 & 1.25$\times10^{-13}$ & 40.59 & 2.96$\times10^{-14}$ & 39.96\\
NGC\,3079 & 2.94$\times10^{-12}$ & 40.90 & 1.23$\times10^{-12}$ & 40.52\\
NGC\,3735 & 9.04$\times10^{-13}$ & 41.15 & 2.47$\times10^{-13}$ & 40.58\\
NGC\,4235 & 1.26$\times10^{-13}$ & 40.20 & 6.25$\times10^{-14}$ & 39.89\\
NGC\,4388 & 2.22$\times10^{-12}$ & 41.48 & 1.59$\times10^{-12}$ & 41.33\\
NGC\,4565 & 1.24$\times10^{-12}$ & 40.64 & 8.35$\times10^{-13}$ & 40.47\\
NGC\,5866 & ... & ... & ... & ... \\
IC\,1368 & 9.04$\times10^{-14}$ & 40.47 & 2.34$\times10^{-14}$ & 39.88\\
UGC\,12282 & 4.18$\times10^{-14}$ & 40.37 & 1.83$\times10^{-14}$ & 40.01\\
\hline
\end{tabular}
\end{flushleft}
Notes: Column~(1) lists the galaxy name. Column~(2) lists the total \ha flux
and column~(3) lists the corresponding luminosity, assuming the distances
quoted in Table~\ref{t:basics}. Similarly, columns~(4) and (5) list the
total \oxy fluxes and luminosities. For NGC\,5866 no accurate flux
measurements were possible, so we did not list any.
\end{table*}
To check our calibration and continuum
subtraction method, we compared our \ha fluxes for five galaxies with
previously published values \citep{col96,leh96b}. The comparison is
summarized in Table~\ref{t:lit}.
\begin{table}
\begin{flushleft}
\caption{Comparison of \ha luminosities with literature values}
\label{t:lit}
\centering
\begin{tabular}{lcc}
\hline\hline
Galaxy & $\log{L}($H$\alpha$) & $\log{L_{\rm{lit}}}$(H$\alpha$) \\
(1) & (2) & (3) \\
\hline
Mrk\,993 & 40.68 & 40.87$^{a}$ \\
Ark\,79 & 40.99 & 40.95$^{a}$ \\
NGC\,3079 & 40.90 & 41.06$^{b}$ \\
NGC\,4235 & 40.20 & 40.38$^{a}$ \\
IC\,1368 & 40.47 & 40.56$^{a}$ \\
\hline
\end{tabular}
\end{flushleft}
Notes: Column~(1) lists the galaxy name. Column~(2) lists our total \ha
luminosity and column~(3) lists the total \ha luminosity quoted in the
literature. ($a$) Data taken from \citet{col96}, ($b$) Data taken from
\citet{leh96b}.
\end{table}
 The published \ha luminosities are in
fairly good agreement with our values, with on average $\log(L_{\rm
us} /L_{\rm lit}) = -0.12\pm0.09$. This is probably a fair assessment
of the accuracy of our absolute flux calibration, including both
random and systematic errors. We did not find any published \oxy
luminosities for any of the galaxies in our sample. However, the
absolute flux calibration of the \ha and \oxy data was carried out in
an analogous fashion. So we expect the uncertainties in both to be
comparable.

\subsubsection{Detection Sensitivity}
\label{ss:sensitivity}

To calculate the sensitivity of our observations for detection of
faint low surface brightness emission we quantified the noise in the
final emission-line maps. We did this using two separate methods.

In the first method we calculated the noise on the basis of
statistical considerations, using the known sky background, detector
properties and observational setup. The values of the sky background
computed for the sky subtraction (see Section~\ref{sss:datared}) were
converted from counts to electrons using the gain values listed on the
BUSCA webpage (see footnote 1); i.e., 1.2136\,$e^{-}$/count for the
$B$ and \oxy images, and 2.9674\,$e^{-}$/count for the $R$ and \ha
images. The result was used to estimate the contribution to the noise
from Poisson statistics in the final broadband and narrowband
images. To this we added in quadrature the contribution from
read-noise, taking into account the number of exposures. The detector
read-noise values are also given on the BUSCA webpage; i.e.,
6.37\,$e^{-}$ for the $B$ and \oxy images, and 7.60\,$e^{-}$ for the
$R$ and \ha images. The results were used to estimate the 1-$\sigma$
noise level in electrons per pixel in each final continuum-subtracted
emission-line image. These were used to calculate the noise in
electrons per square arcsec, using the known pixel size and the fact
that noise from co-added independent pixels adds in quadrature. The
results were transformed to units of
ergs\,s$^{-1}$\,cm$^{-2}$\,arcsec$^{-2}$ using the detector gain
values and the flux calibrations described in Section~\ref{ss:flux}.
The predicted 1-$\sigma$ noise values for the \ha and \oxy
emission-line maps are listed in Table~\ref{t:sens}.
\onltab{5}{
\begin{table}
\begin{flushleft}
\caption{Sensitivity limits for the \ha and \oxy emission-line maps}
\label{t:sens}
\centering
\begin{tabular}{lcccccc}
\hline\hline
 & \multicolumn{2}{c}{Predicted 1-$\sigma$ noise} & \multicolumn{2}{c}{Measured
1-$\sigma$ noise} & \multicolumn{2}{c}{Predicted 3-$\sigma$ noise}\\
Galaxy & \ha & \oxy & \ha & \oxy & \ha & \oxy \\
&  \multicolumn{2}{c}{[$\rm{ergs\,s^{-1}\,cm^{-2}\,arcsec^{-2}}$]} &
\multicolumn{2}{c}{[$\rm{ergs\,s^{-1}\,cm^{-2}\,arcsec^{-2}}$]} &
\multicolumn{2}{c}{[$\rm{ergs\,s^{-1}\,cm^{-2}\,arcsec^{-2}}$]} \\
(1) & (2) & (3) & (4) & (5) & (6) & (7)\\
\hline
Mrk\,993 & $1.14\times10^{-17}$ & $2.56\times10^{-17}$ & $1.03\times10^{-17}$ &
$2.97\times10^{-17}$ & $3.41\times10^{-17}$ & $7.69\times10^{-17}$  \\
Mrk\,577 & $1.03\times 10^{-17}$ & $2.20\times 10^{-17}$ & $8.62\times 10^{-18}$ &
$2.72\times10^{-17}$ & $3.10\times10^{-17}$   &  $6.59\times 10^{-17}$  \\
UGC\,1479 & $8.61\times10^{-18}$ & $1.73\times10^{-17}$ & $8.21\times10^{-18}$ &
$2.07\times10^{-17}$ & $2.58\times10^{-17}$ & $5.18\times10^{-17}$  \\
Ark\,79 & $1.25\times10^{-17}$ & $2.71\times10^{-17}$ & $1.10\times10^{-17}$ &
$2.98\times10^{-17}$ & $3.74\times10^{-17}$ & $8.12\times10^{-17}$  \\
Mrk\,1040 & $1.20\times10^{-17}$ & $2.48\times10^{-17}$ & $1.17\times10^{-17}$ &
$2.68\times10^{-17}$ & $3.61\times10^{-17}$ & $7.44\times10^{-17}$  \\
UGC\,2936 & $1.28\times10^{-17}$ & $2.71\times10^{-17}$ & $1.19\times10^{-17}$ &
$3.05\times10^{-17}$ & $3.85\times10^{-17}$ & $8.13\times10^{-17}$  \\
NGC\,3079 & $1.20\times10^{-17}$ & $1.72\times10^{-17}$ & $1.22\times10^{-17}$ &
$2.12\times10^{-17}$ & $3.59\times10^{-17}$ & $5.16\times10^{-17}$  \\
NGC\,3735 & $2.44\times10^{-17}$ & $2.97\times10^{-17}$ & $2.19\times10^{-17}$ &
$3.29\times10^{-17}$ & $7.32\times10^{-17}$ & $8.92\times10^{-17}$  \\
NGC\,4235 & $1.51\times10^{-17}$ & $2.39\times10^{-17}$ & $1.21\times10^{-17}$ &
$2.66\times10^{-17}$ & $4.54\times10^{-17}$ & $7.17\times10^{-17}$  \\
NGC\,4388 & $3.10\times10^{-17}$ & $4.43\times10^{-17}$ & $2.32\times10^{-17}$ &
$4.09\times10^{-17}$ & $9.31\times10^{-17}$ & $1.33\times10^{-16}$  \\
NGC\,4565 & $1.13\times10^{-17}$ & $1.74\times10^{-17}$ & $9.94\times10^{-18}$ &
$2.21\times10^{-17}$ & $3.39\times10^{-17}$ & $5.21\times10^{-17}$  \\
NGC\,5866 & $1.23\times10^{-17}$ & $1.92\times10^{-17}$ & $1.06\times10^{-17}$ &
$2.19\times10^{-17}$ & $3.68\times10^{-17}$ & $5.76\times10^{-17}$  \\
IC\,1368 & $9.09\times10^{-18}$ & $1.81\times10^{-17}$ & $8.44\times10^{-18}$ &
$2.38\times10^{-17}$ & $2.73\times10^{-17}$ & $5.42\times10^{-17}$  \\
UGC\,12282 & $8.56\times10^{-18}$ & $1.59\times10^{-17}$ & $7.74\times10^{-18}$ &
$1.96\times10^{-17}$ & $2.57\times10^{-17}$ & $4.78\times10^{-17}$  \\
\hline
\end{tabular}
\end{flushleft}
Notes: Column~(1) lists the galaxy name. Columns~(2) and (3) list the 1-$\sigma$ noise
levels found from computing the total noise arising from the Poisson noise of the sky
background and the readout noise of the CCD chips. Columns~(4) and (5) list the
1-$\sigma$ noise levels measured in the emission-line maps. Columns~(6) and (7) list
the predicted 3-$\sigma$ noise level (i.e. columns~(2) and (3) multiplied by 3).
\end{table}
}

In the second method for estimating the noise in the emission-line
maps we used the histogram of the pixel values in an ``empty'' region
of each emission-line map (i.e., away from the galaxy, bright stars,
ghosts, and detector artifacts). We measured the FWHM and divided by
$2.355$ to obtain an estimate of the 1-$\sigma$ noise level in counts
per pixel (a Gaussian distribution has $\sigma = {\rm FWHM}/2.355$;
basing the calculation on the FWHM value rather than the second moment
has the advantage of robustness against outlier pixels).  This was
transformed to units of ergs\,s$^{-1}$\,cm$^{-2}$\,arcsec$^{-2}$ in
the same manner as for the first method. The measured 1-$\sigma$ noise
values for the \ha and \oxy emission-line maps are also listed in
Table~\ref{t:sens}. The second method has the advantage that it takes
into account potential additional sources of noise that cannot be
accounted for by idealized statistical calculations (e.g.,
flat-fielding errors). However, the predicted and measured values are
overall in very good agreement, which gives extra confidence in the
results.

The \ha and \oxy narrow-band observations contribute roughly twice as
much noise to the final emission-line maps as the scaled $R$ and
$B$-band images. The main contributor to the noise in the \ha and \oxy
narrow-band images is the Poisson noise from the background, which is
on average two to four times as large as the contribution from the
readout noise. The galaxy with the lowest predicted and measured noise
level in both the \ha and \oxy emission line maps is UGC\,12282, with
a predicted 1-$\sigma$ noise level of
$8.56\times10^{-18}$\,ergs\,s$^{-1}$\,cm$^{-2}$\,arcsec$^{-2}$ in \ha
and $1.59\times10^{-17}$\,ergs\,s$^{-1}$\,cm$^{-2}$\,arcsec$^{-2}$ in
\oxy. The galaxy with the highest noise level is NGC\,4388, since for
this galaxy only one frame was available in each filter. The predicted
1-$\sigma$ noise levels for this galaxy are
$3.10\times10^{-17}$\,ergs\,s$^{-1}$\,cm$^{-2}$\,arcsec$^{-2}$ in \ha
and $4.43\times10^{-17}$\,ergs\,s$^{-1}$\,cm$^{-2}$\,arcsec$^{-2}$ in
\oxy.

We take three times the predicted 1-$\sigma$ noise level for each
emission line map as a measure of our final sensitivity for detection
of faint low surface brightness emission. The 3-$\sigma$ detection
sensitivities thus obtained for each galaxy are also listed in
Table~\ref{t:sens}. The median sensitivity over the sample is
$3.6\times10^{-17}$\,ergs\,s$^{-1}$\,cm$^{-2}$\,arcsec$^{-2}$ for the
\ha images and
$6.9\times10^{-17}$\,ergs\,s$^{-1}$\,cm$^{-2}$\,arcsec$^{-2}$ for the
\oxy images. The \ha intensity $I$ can be directly transformed into a
line emission measure (EM), which is defined as the line-of-sight
integral over the squared electron density. For an ionized region that
is thick to Lyman lines (case B recombination) one has \citep{rey92}
that
\begin{equation}
  \frac{\rm EM}{{\rm cm}^{-6}\,\pc} \approx 
    \left(\frac{I}{2.0 \times 10^{-18}\,{\rm ergs}\,{\rm s}^{-1}\,{\rm cm}^{-2}\,{\rm arcsec}^{-2}}\right) 
    \left(\frac{T}{10^4 {\rm K}}\right)^{-0.92}.
\end{equation}
So if we assume an electron temperature $T \approx 8000$ K as in
\citet{vei95}, then our median sensitivity limit for \ha corresponds
to $22 \,{\rm cm}^{-6}\,\pc$.

The quoted sensitivities refer to detection over an area of 1 square
arcsec. If one averages over larger areas then one can potentially see
fainter emission, but there will be less information on spatial
detail. Point-source sensitivities (which are of less interest in the
context of the present paper) are similar to the values listed in
Table~\ref{t:sens}, given that the seeing FWHM of most observations
was close to $1$ arcsec (see Table~\ref{t:obs}).  The quoted limits do
not take into account the noise and subtraction problems associated
with galaxy continuum light. Given the edge-on viewing of the
galaxies, the limits therefore apply to high-latitude emission-line
gas only.

Our sensitivity limits are comparable to those determined for
NGC\,3079 by \citet{vei94,vei95}, who quote
$\sim3\times10^{-17}$\,ergs\,s$^{-1}$\,cm$^{-2}$\, arcsec$^{-2}$. Our
sensitivities are also similar to those achieved in the study of 9
galaxies by \citet{ran96}, who obtained observations with a comparable
instrumental setup. Of course, pointed studies of individual galaxies
with 8m-class telescopes \citep[e.g.,][]{yos02} or detailed
Fabry-Perot studies on 4m-class telescopes \citep[e.g.,][]{vei03} can
achieve higher sensitivities than those obtained here. For comparison,
for NGC\,4388 our observations are a factor of four and ten (\oxy and
\ha, respectively) less sensitive than the very deep SUBARU
observations by \citet{yos02}. (But our \ha sensitivity for this
galaxy is low by a factor $2.6$ relative to the median for the other
galaxies in our own sample, due to the fact that we had only one
exposure available.)

\subsubsection{The ionization maps}
\label{ss:imaps}

Finally, \oxy/\ha ionization maps were constructed using the following 
procedure. First, the average remaining background value was subtracted from 
each \ha and \oxy map. This was done since in some cases the background value
was not exactly zero, due to the process of combining the broadband and 
narrowband frames. We then performed a basic $1\,\sigma$ clipping, by setting 
all pixels with count values lower than $1\,\sigma$ of the background to 
zero. This included any negative count values. This was necessary in order 
to avoid divisions by very small numbers found in noise, which if divided by, 
would lead to values of the \oxy/\ha ratio larger than the actual signal 
itself. Following this, we divided the \oxy image by the \ha image in order 
to obtain the final ionization maps. We then applied a $3\times3$ median 
box filter to the ionization maps in order to reduce the noise. This approach 
is similar to the method used by \citet{pog88a}. In 
Section~\ref{ss:imapdiscuss} we discuss the results of the ionization maps, 
presented in Section~\ref{s:res}.

\section{Results}
\label{s:res}

In the following descriptions, we quote angular separations and sizes of 
interest, and in each case we quote in brackets the corresponding projected 
physical sizes, assuming the distances given in Table \ref{t:basics}.

\subsection{Mrk\,993}

\begin{figure*}
\centering
\includegraphics[scale=0.8]{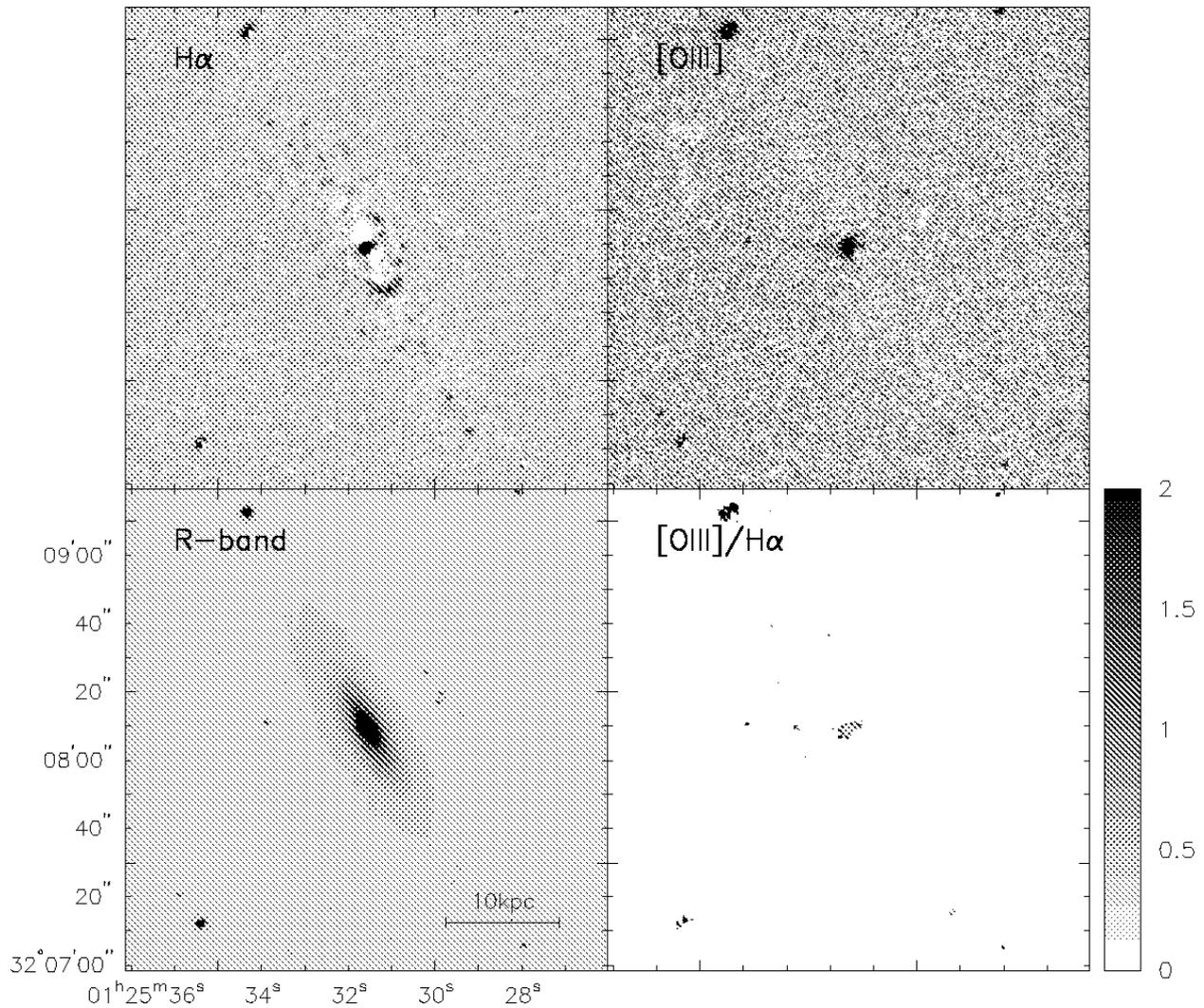}
\caption{Broadband, narrowband images and ionization map of Mrk\,993. \ha
(upper left), \oxy (upper right), $R$-band (lower left), and \oxy/\ha (lower
right). \label{f:mrk993}}
\end{figure*}

Mrk\,993 (Figure~\ref{f:mrk993})
 appears to be a relatively featureless 
spiral in the $R$-band. The \ha map shows emission from the nucleus, 
\hbox{H\,{\sc ii}} regions scattered throughout the disk (as previously 
observed by \citealt{col96}), and traces of diffuse emission between the 
compact emission regions. The \hbox{H\,{\sc ii}} regions are visible out to 
$53\arcsec$ (16.0~kpc) to the NE and $64\arcsec$ (19.3~kpc) to the SW of the 
nucleus. Also visible is an \ha emitting ring-like structure of radius 
$\sim14\arcsec$ ($\sim4.2$~kpc) around the nucleus. Most of the emission 
visible in the \oxy map originates from the nucleus of the galaxy, and none 
of the widespread emission seen in \ha is seen in \oxy. No extraplanar \ha 
emission is observed, in agreement with \citet{col96}, and no extraplanar 
\oxy emission is observed.

\subsection{Mrk\,577}

The disk of Mrk\,577 (Figure~\ref{f:mrk577})
\onlfig{6}{
\begin{figure*}
\centering
\includegraphics[scale=0.8]{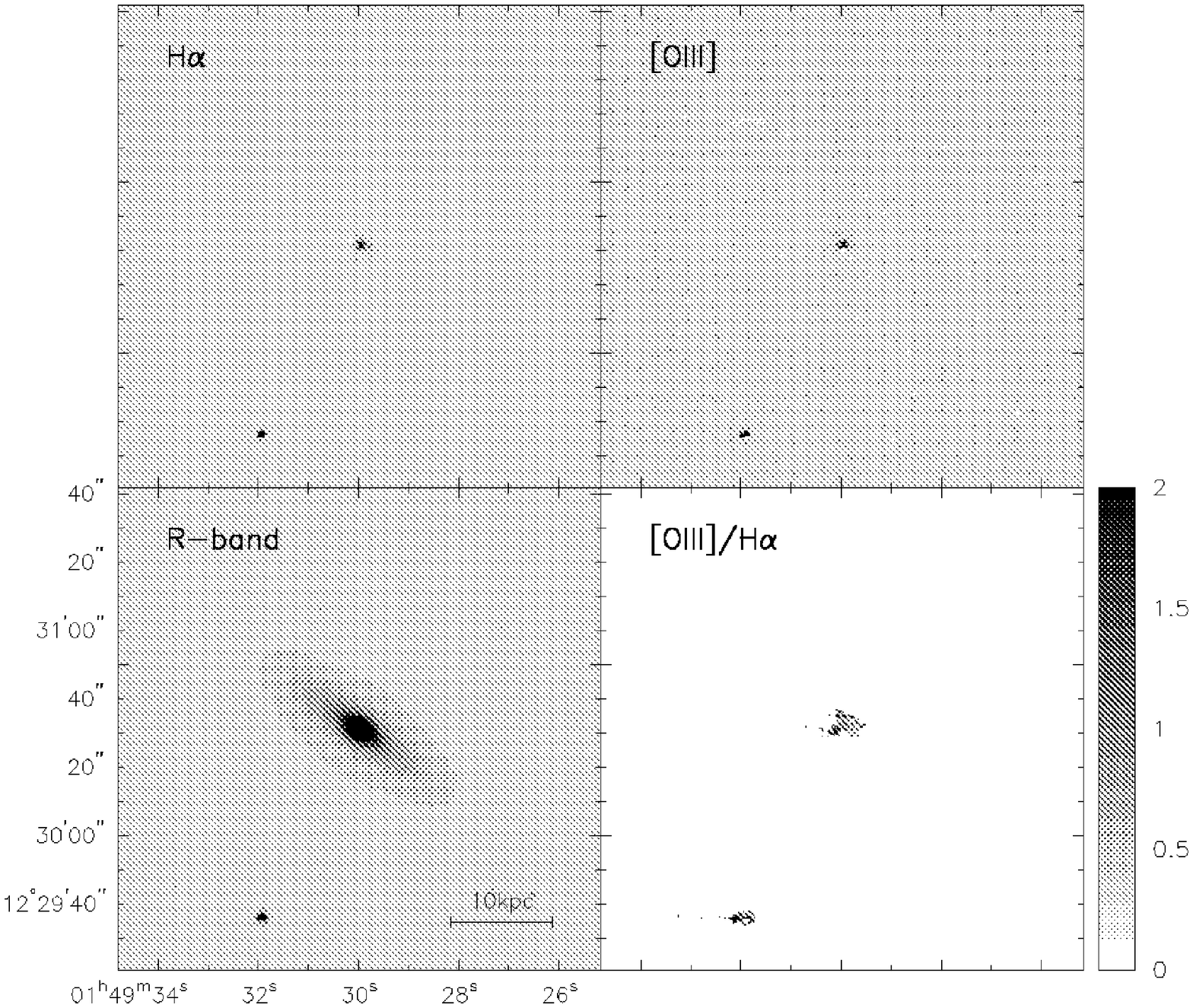}
\caption{Broadband, narrowband images and ionization map of Mrk\,577. \ha
(upper left), \oxy (upper right), $R$-band (lower left), and \oxy/\ha (lower
right). \label{f:mrk577}}
\end{figure*}
}
is remarkably featureless in the 
$R$-band, and only the nucleus is visible 
in the \ha and the \oxy maps. The nucleus in the \ha map 
is slightly offset in the NW direction compared to the brightest peak in the 
$R$-band. This may be due to dust extinction in the circumnuclear 
environment. No planar or extraplanar \ha or \oxy emission is observed for 
this galaxy. Therefore, only the nucleus is present in the ionization map.

\subsection{UGC\,1479}

The $R$-band image of UGC\,1479 (Figure~\ref{f:ugc1479})
\onlfig{7}{
\begin{figure*}
\centering
\includegraphics[scale=0.8]{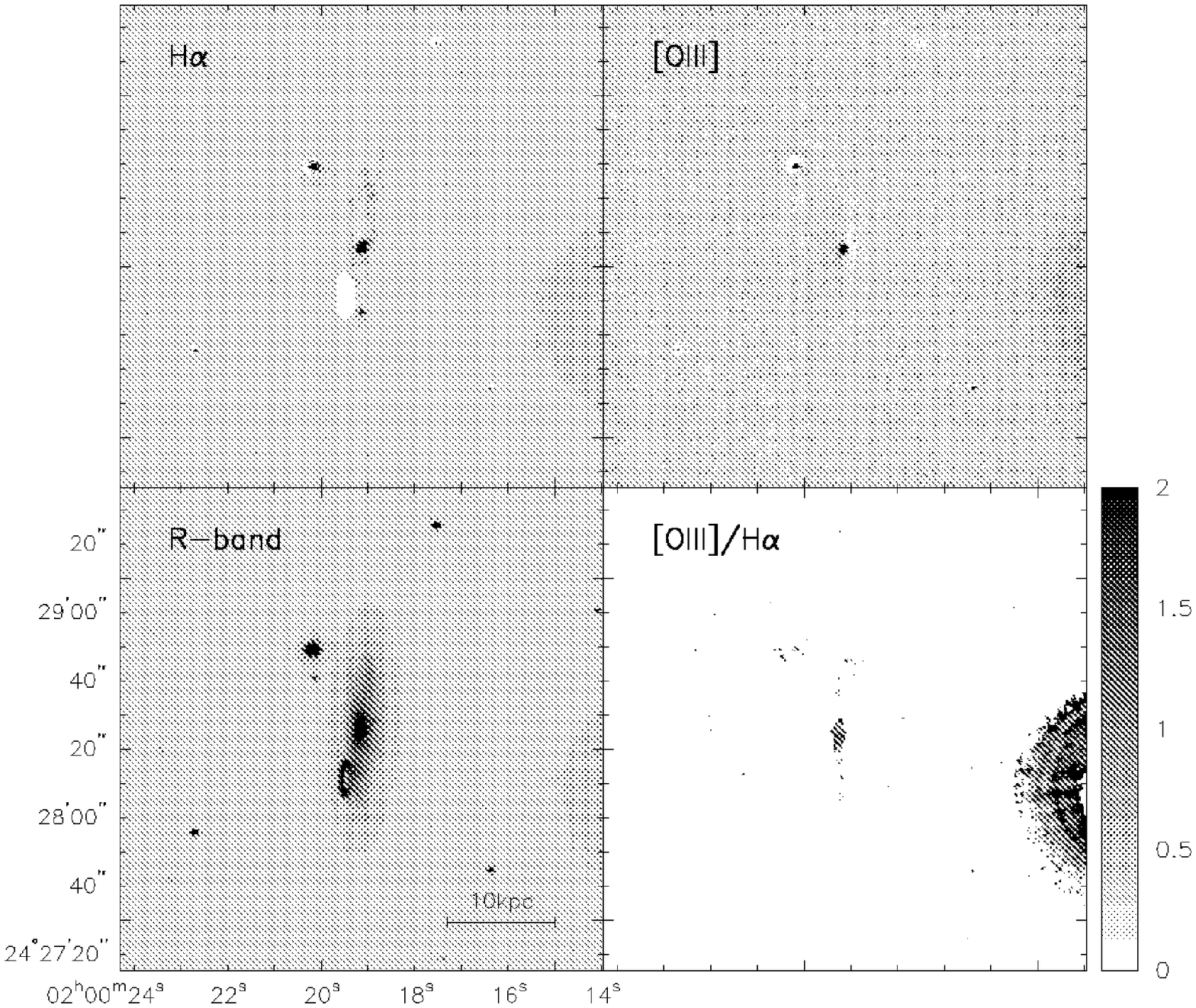}
\caption{Broadband, narrowband images and ionization map of UGC\,1479. \ha
(upper left), \oxy (upper right), $R$-band (lower left), and \oxy/\ha (lower
right). The oval white region in the \ha map and the corresponding feature in
the $R$-band image close to the southern part of the disk is a ghost of a
bright star just off to the right of the displayed field of view (see
Section~\ref{ss:artifacts}). \label{f:ugc1479}}
\end{figure*}
}
 appears relatively 
featureless apart from the prominent nuclear region. The \ha map clearly 
shows emission from two spiral arm-like structures, as well as extended 
emission from the nucleus. The morphology of the latter is reminiscent of a 
bar joining the base of the two spiral arms. Diffuse \ha emission is observed 
only within the disk, and extends $32\arcsec$ (10.2~kpc) to the N and 
$25\arcsec$ (8.0~kpc) to the S of the nucleus. The \oxy map also reveals 
emission from the nucleus, and faint diffuse emission from the disk, which 
seems offset to the E, possibly due to inclination effects. However, this 
diffuse emission may also be due to residuals from the continuum subtraction. 
No extraplanar emission is seen in either \ha or \oxy. The black oval region 
seen in the $R$-band frame is a reflection from a bright star (see Section 
\ref{ss:artifacts}). It appears as a white oval in the \ha image due to the
continuum subtraction process.

\subsection{Ark\,79}

The $R$-band image of Ark\,79 (Figure~\ref{f:ark79})
\onlfig{8}{
\begin{figure*}
\centering
\includegraphics[scale=0.8]{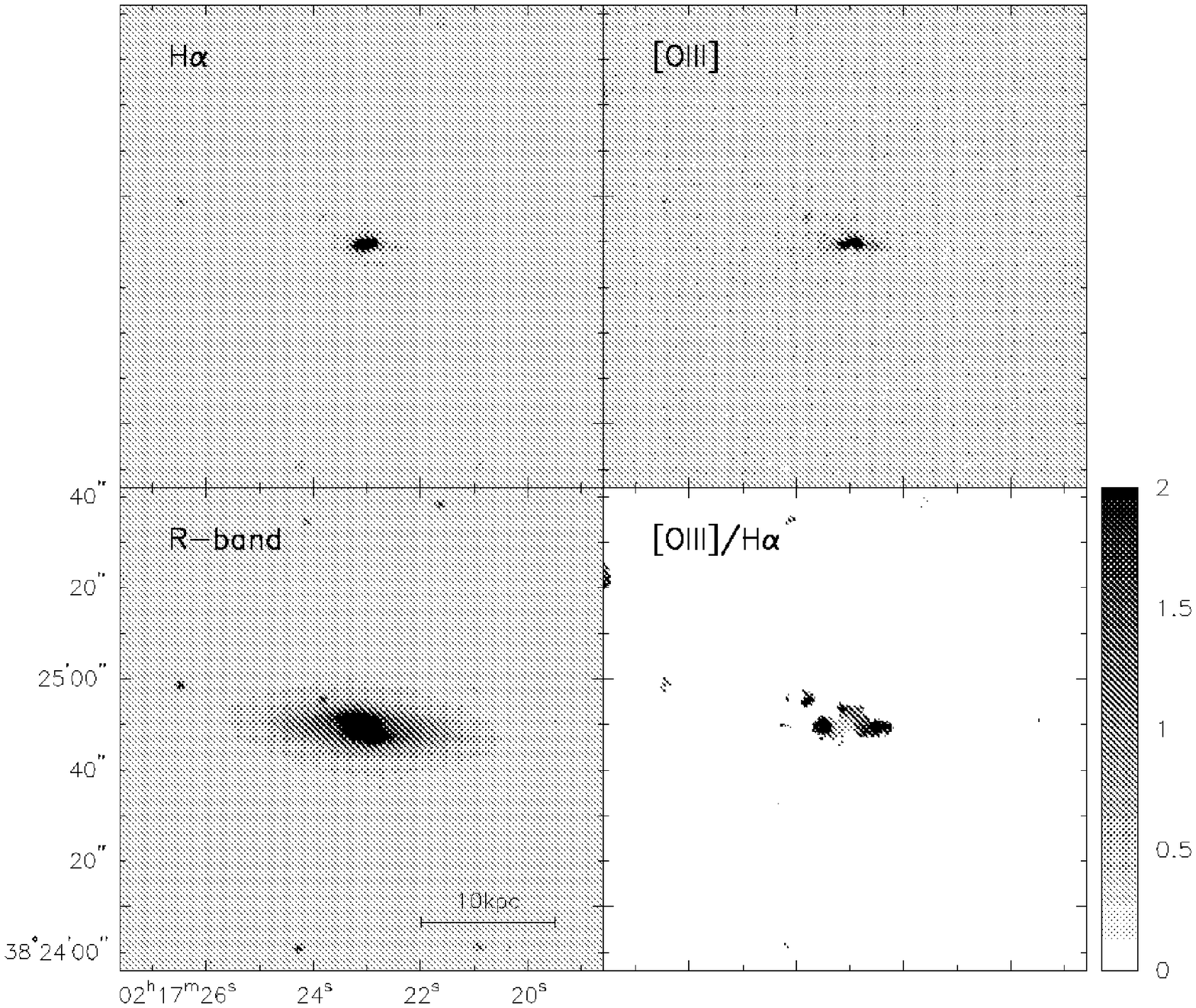}
\caption{Broadband, narrowband images and ionization map of Ark\,79. \ha
(upper left), \oxy (upper right), $R$-band (lower left), and \oxy/\ha (lower
right). \label{f:ark79}}
\end{figure*}
}
 shows one nucleus and a 
structure which may be a faint inner bar. The nucleus is the main 
contributor to the emission in the \ha and \oxy maps. When shown on a 
suitable greyscale, the narrowband maps reveal two components to the nucleus 
(see a zoomed-in version of the nuclear region in Figure~\ref{f:ark79z}).
\onlfig{9}{
\begin{figure*}
\centering
\includegraphics[scale=0.8]{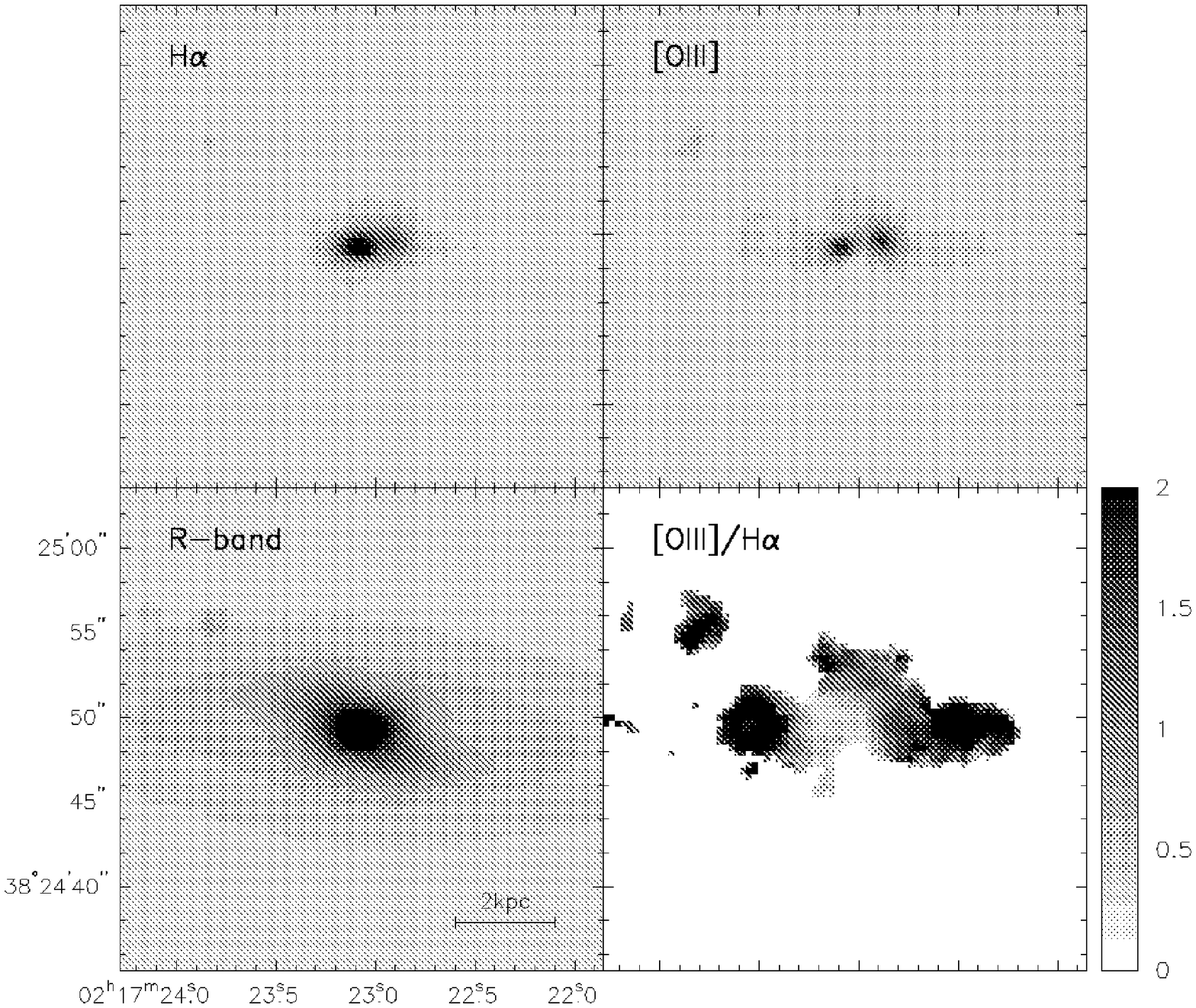}
\caption{A blow-up version of Figure~\ref{f:ark79}, showing the nuclear
region of Ark\,79. The \ha and \oxy images are also shown in a slightly
different greyscale stretch, to highlight the two components more clearly.
\label{f:ark79z}}
\end{figure*}
}
 The 
brightest peak in \ha is located to the E, centered on the broadband peak, 
and the fainter one is offset to the W. However, the two components have a 
similar brightness in \oxy, meaning that the \oxy/\ha ionization ratio will 
be much larger for the offset peak compared to the actual center of the 
galaxy. The separation between the two components is $2\farcs5$ (850\,pc).
This is the first reported detection of a two component nucleus in 
Ark\,79. \ha and \oxy emission is also seen extending into the disk, and
resembles the morphology observed by \citet{col96}. No clear extraplanar
emission is visible on large scales, also in agreement with \cite{col96}.

\subsection{Mrk\,1040}

Our broadband image (see Figure~\ref{f:mrk1040})
\onlfig{10}{
\begin{figure*}
\centering
\includegraphics[scale=0.8]{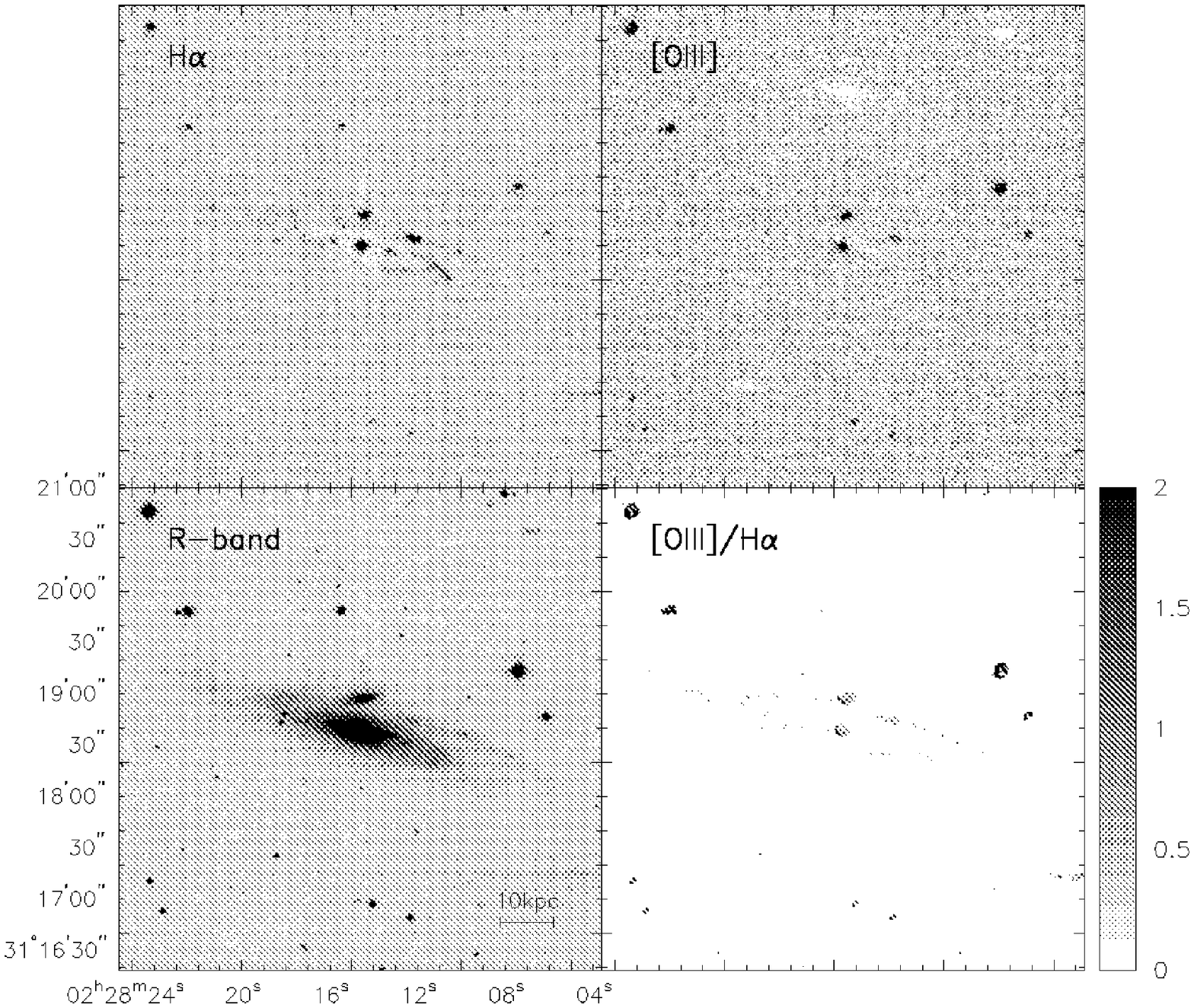}
\caption{Broadband, narrowband images and ionization map of Mrk\,1040. \ha
(upper left), \oxy (upper right), $R$-band (lower left), and \oxy/\ha (lower
right). \label{f:mrk1040}}
\end{figure*}
}
 shows the presence of a 
companion galaxy $18\arcsec$ to the 
N of the nucleus of Mrk\,1040, which may be in the process of merging. This
was also previously reported by 
\citet{col96}. The \ha emission from the \hbox{H\,{\sc ii}} regions in the 
disk of Mrk\,1040 outlines a complex spiral arm structure, with bright 
\hbox{H\,{\sc ii}} regions seen as far out as $119\arcsec$ (38.4~kpc) to the 
E of the nucleus, and $86\arcsec$ (27.8~kpc) to the W. This asymmetry 
could be explained by interactions with other galaxies such as the northern 
companion. The nucleus of Mrk\,1040 and that of the companion galaxy, as well 
as one region in the spiral arms $32\arcsec$ (10.3~kpc) to the W of the 
nucleus of Mrk\,1040 are strong \ha and \oxy emitters (a closeup view is 
shown in Figure~\ref{f:mrk1040z}),
\onlfig{11}{
\begin{figure*}
\centering
\includegraphics[scale=0.8]{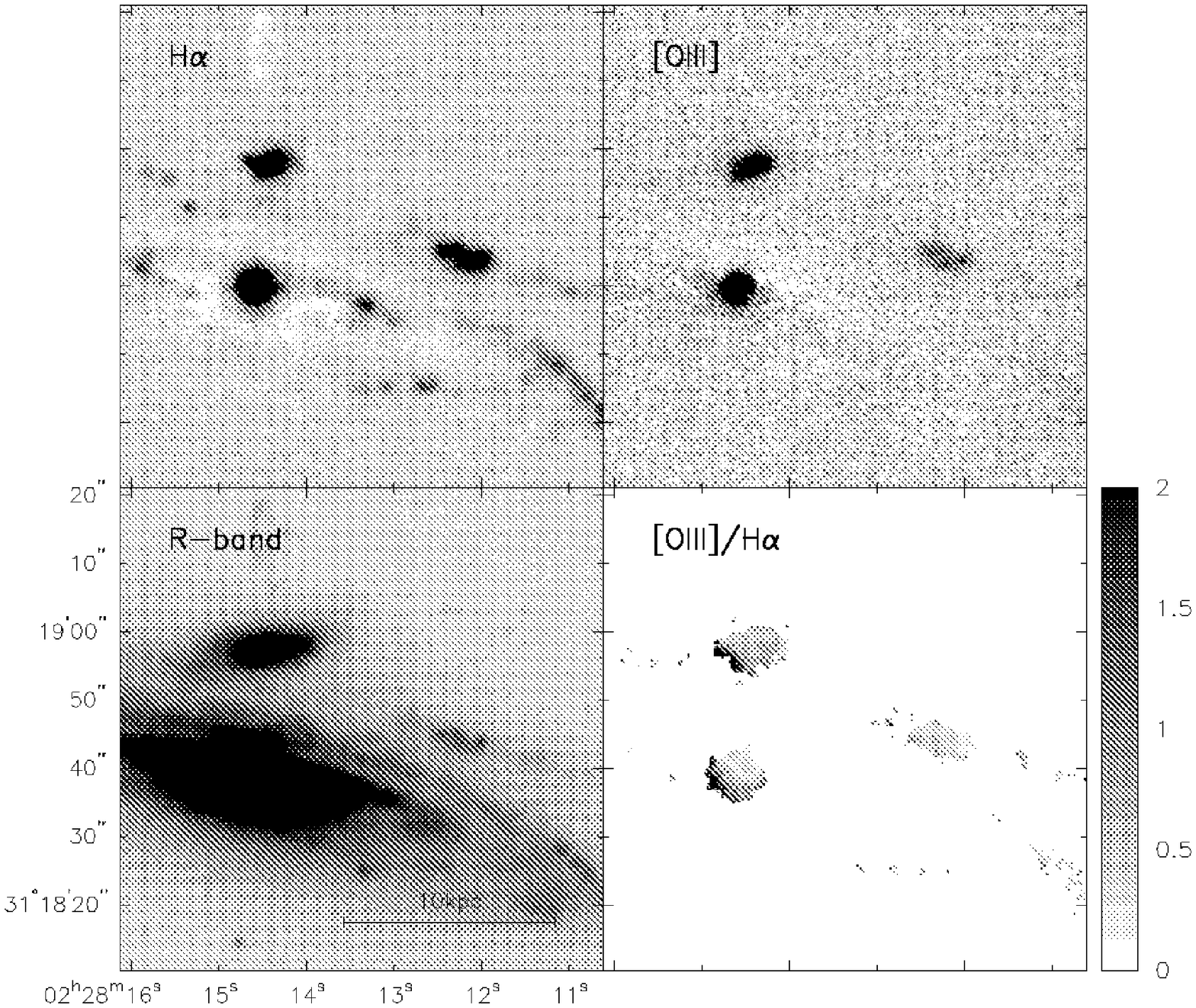}
\caption{A zoom-in of Figure~\ref{f:mrk1040}, highlighting the central
parts of Mrk\,1040.\label{f:mrk1040z}}
\end{figure*}
}
 and there is a possible a ring-like 
structure of radius $20\arcsec$ (6.5~kpc) around the nucleus in \ha. The 
remaining fraction of the \oxy emission originates from faint diffuse 
emission in the spiral arms. No evidence for extraplanar emission is found in 
either \ha or \oxy. The ionization map shows mainly the two galactic nuclei 
and the bright \hbox{H\,{\sc ii}} region within the spiral arm.

\subsection{UGC\,2936}

Our $R$-band image of UGC\,2936 (Figure~\ref{f:ugc2936})
\onlfig{12}{
\begin{figure*}
\centering
\includegraphics[scale=0.8]{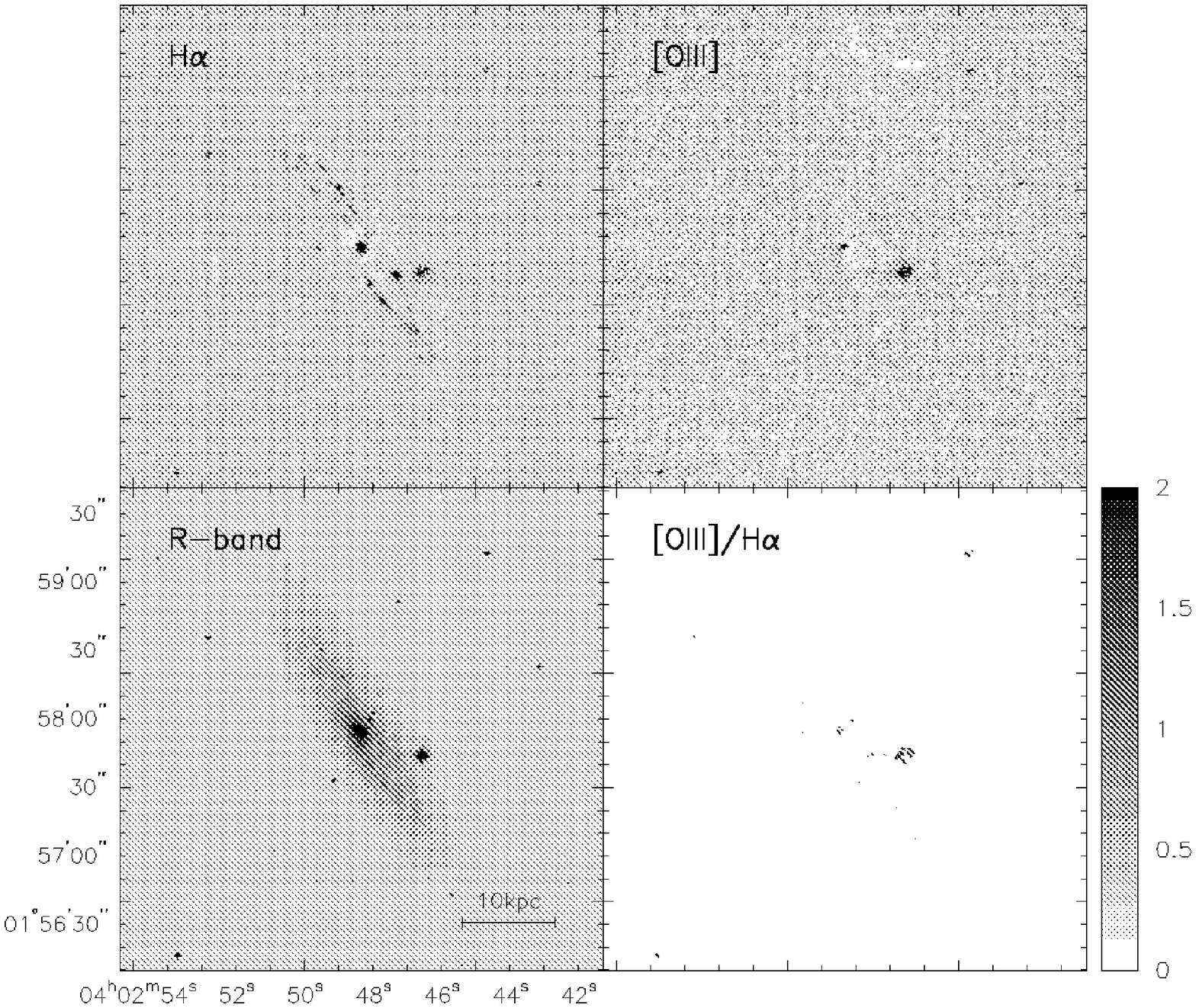}
\caption{Broadband, narrowband images and ionization map of UGC\,2936. \ha
(upper left), \oxy (upper right), $R$-band (lower left), and \oxy/\ha (lower
right). \label{f:ugc2936}}
\end{figure*}
}
 reveals the nucleus 
and \hbox{H\,{\sc ii}} regions along spiral arms. These are clearly visible 
in the \ha map, and are seen scattered throughout
the disk out to $68\arcsec$ (16.7~kpc) to the NE and $58\arcsec$ 
(14.3~kpc) to the SW of the nucleus. A few faint clustered 
\hbox{H\,{\sc ii}} regions are visible beyond what seems to be the rim of the 
outer spiral arm. The \oxy map is remarkably featureless, apart from emission 
from the nucleus.

\subsection{NGC\,3079}

Many dusty regions and strong \hbox{H\,{\sc ii}} regions are visible in the 
$R$-band image of NGC\,3079 (Figure~\ref{f:ngc3079}).
\onlfig{13}{
\begin{figure*}
\centering
\includegraphics[scale=0.8]{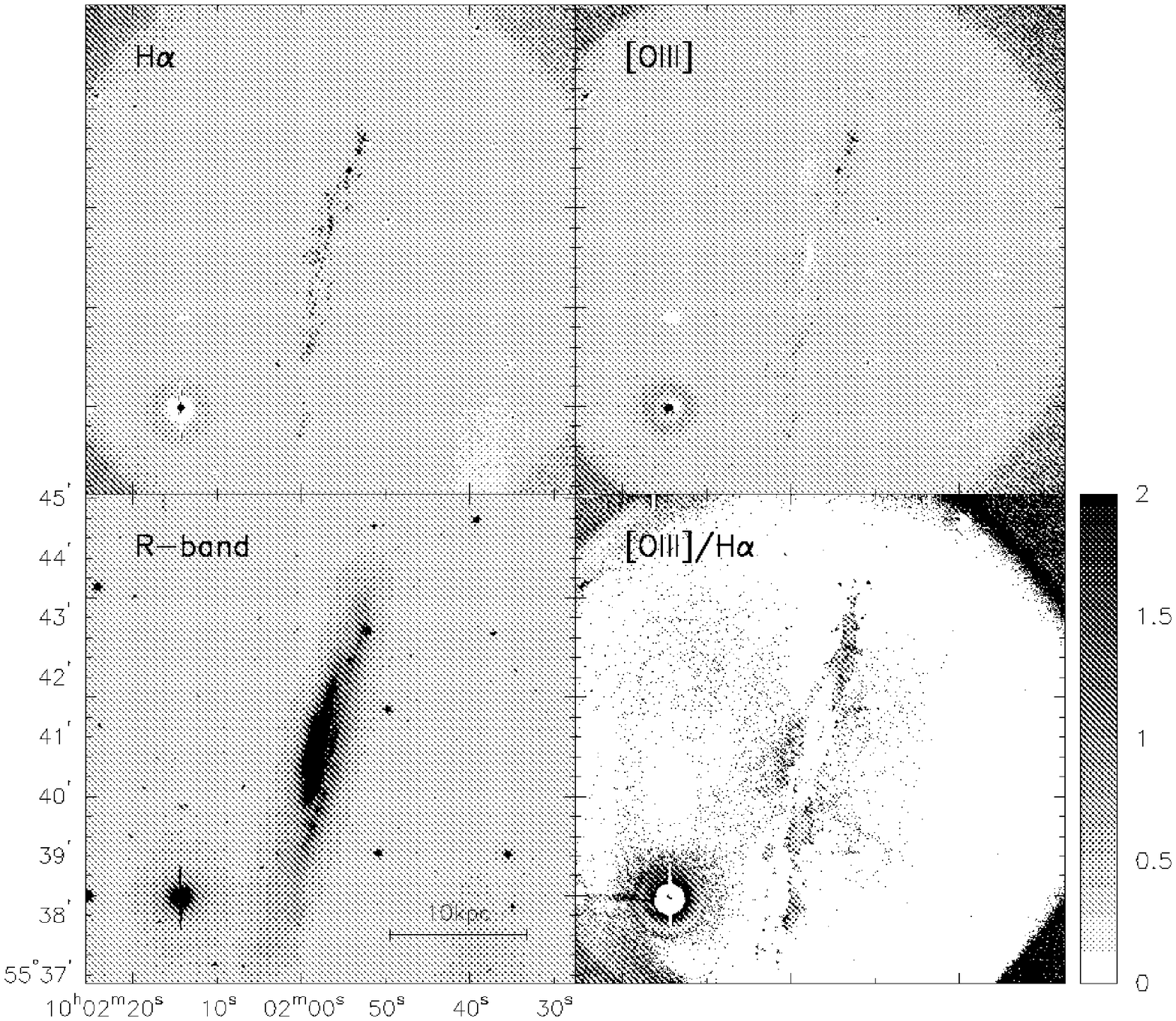}
\caption{Broadband, narrowband images and ionization map of NGC\,3079. \ha
(upper left), \oxy (upper right), $R$-band (lower left), and \oxy/\ha (lower
right). \label{f:ngc3079}}
\end{figure*}
}
 Our \ha image shows 
emission-line regions scattered throughout a warped disk, as well as a 
superbubble in the nuclear region. Both the superbubble 
\citep[e.g.,][]{vei94,cec01,cec02} and the fueling of active nucleus 
\citep[e.g.,][]{vei99b} have been the subject of several previous studies. An 
enlargement of the inner central region of NGC\,3079, tracing the superbubble 
and the outflow cone, is shown in Figure~\ref{f:ngc3079z}.
\onlfig{14}{
\begin{figure*}
\centering
\includegraphics[scale=0.8]{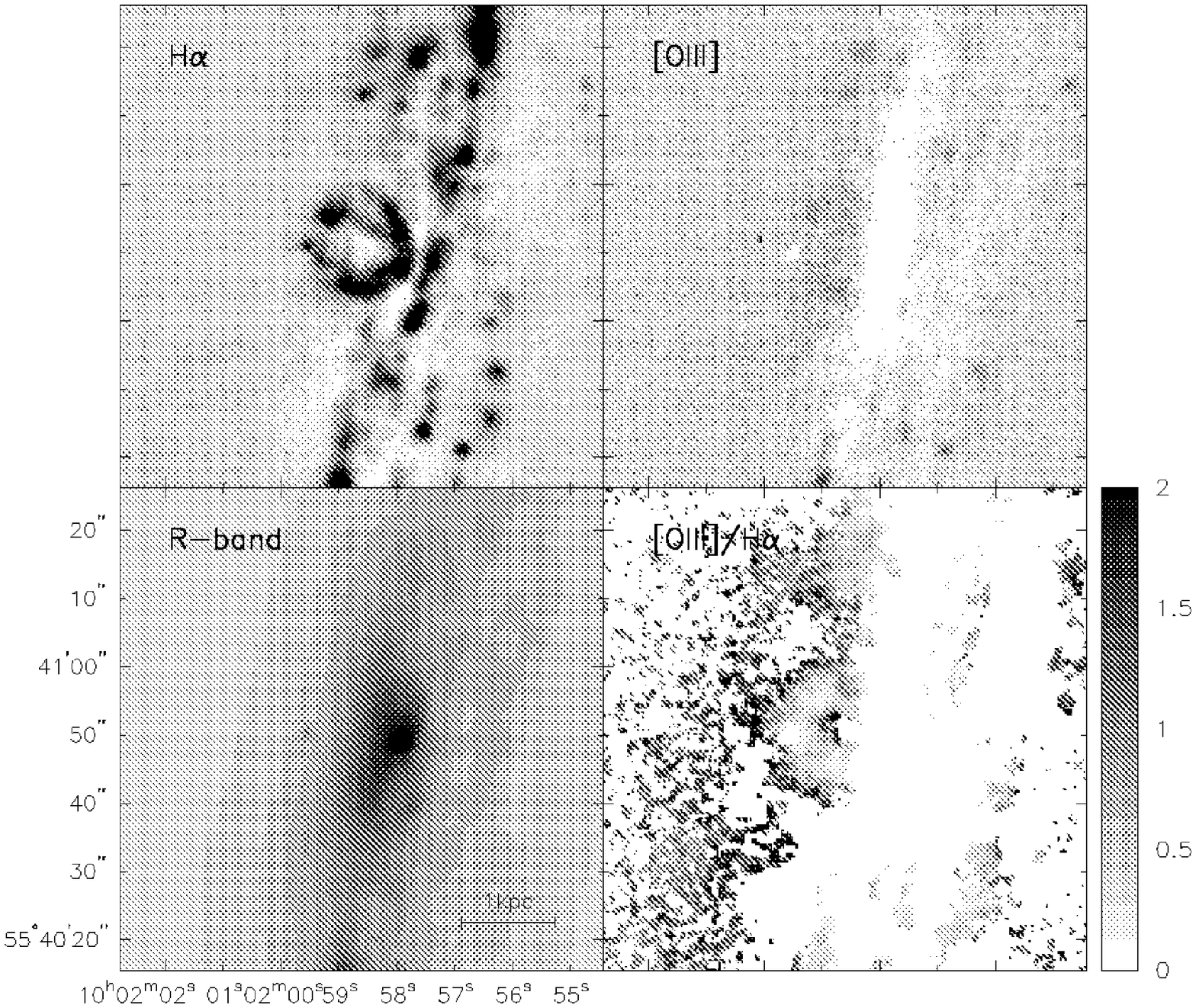}
\caption{A zoom-in of Figure~\ref{f:ngc3079}, showing the central part
(including the superbubble) of NGC\,3079.\label{f:ngc3079z}}
\end{figure*}
}
The fainter extraplanar gas and large-scale outflows emanating from the disk 
are shown in an enhanced contrast \ha map (see Figure~\ref{f:fourha}).
\onlfig{15}{
\begin{figure*}
\centering
\includegraphics[scale=0.8]{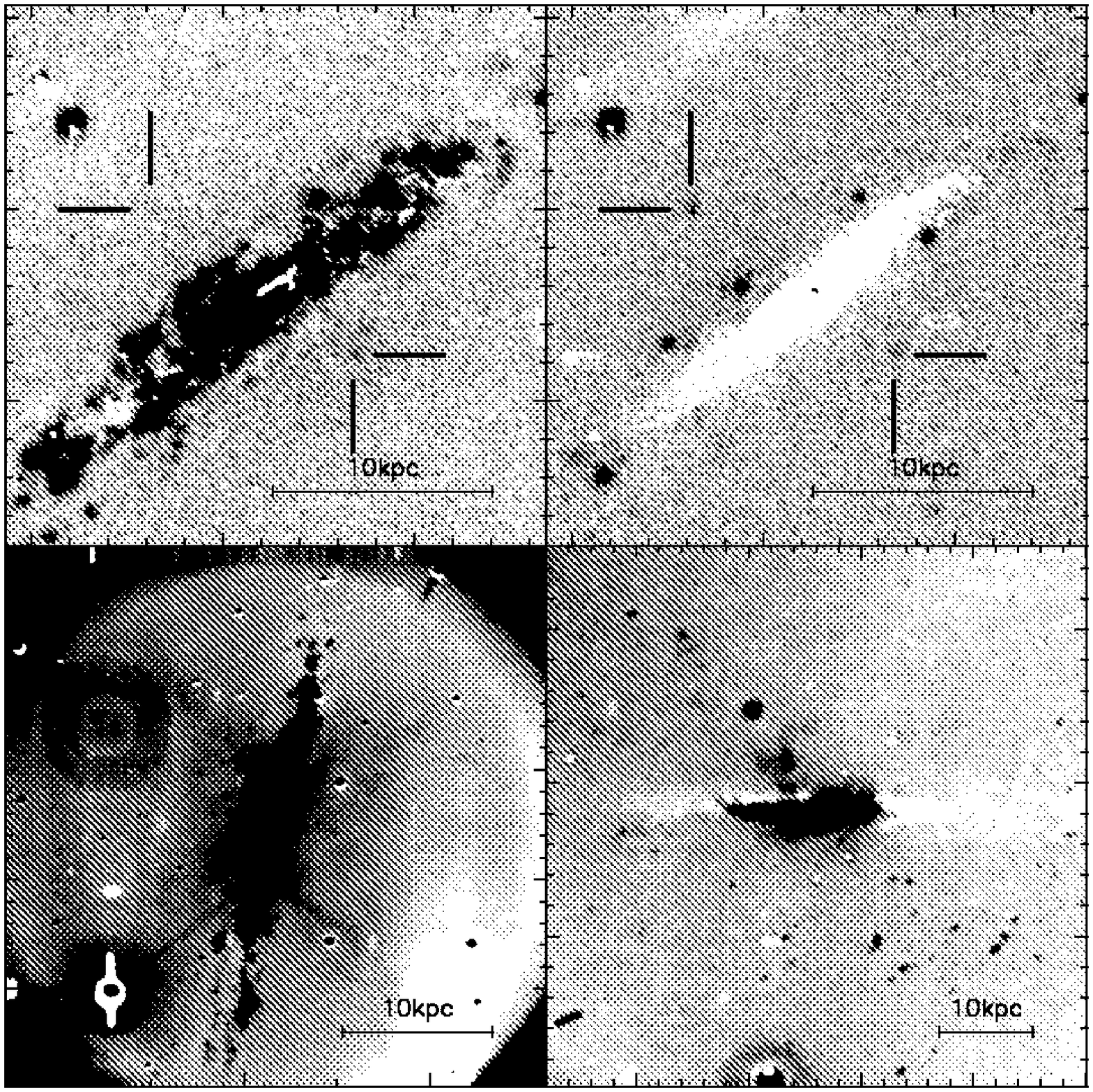}
\caption{High contrast images of NGC\,3079 in \ha (lower left), NGC\,3735
in \ha (upper left), NGC\,3735 in \oxy (upper right) and NGC\,4388 in \ha
(lower right). Note, that there are some remaining cosmic-ray events
(the dark spots mostly to the S of the galaxy) in the \ha image of
NGC\,4388, since only one single image was obtained for this galaxy.
\label{f:fourha}}
\end{figure*}
}
 It is 
interesting to note that all the gaseous outflows and extraplanar diffuse 
emission are only observed to originate within $\sim70''$ (5.1\,kpc) on 
either side of the nucleus. Beyond this radius, the \ha emission simply shows 
bright \ha emission from clumps within the disk, and no extraplanar diffuse 
emission.

The large-scale outflows extend as far out as $\sim185''$ (for the NE 
filament) from the disk midplane ($\sim13.5$kpc). The large-scale filaments 
point radially away from regions in the disk of the galaxy rather than away 
from the nucleus, and seem to trace an X-structure. This was previously 
observed by \cite{hec90} in \ha (although in a smaller field of 
view), and by \cite{str04} in X-ray. Such a morphology is reminiscent of 
the X-structure observed in NGC\,5775 \citep[e.g.,][]{ros03b} and NGC\,4666 
\citep{dah97}, which are starburst galaxies. Out \ha image shows the most 
extended emission, and is the deepest and largest field of view image of
NGC\,3079 to date. The SW outflow seems well collimated, and extends out to 
$\sim125''$ (9.1~kpc). However, we note, that the feature, resembling an 
outflow to the SE (between the galaxy and the bright star) is a 
diffraction spike of the stellar image (see also bottom right panel of 
Figure~\ref{f:fourart}).

The \oxy image shows emission from compact regions throughout the disk, with 
the strongest emission occurring to the N of the nucleus. This region of high 
\oxy emission seems to be close to the base of one of the disk outflows. The 
brightest \oxy regions are not associated with the nucleus, which is most
likely due to heavy dust extinction close to the nucleus.

\subsection{NGC\,3735}

Strong emission from many compact regions in the disk, diffuse emission 
between these regions, and emission from an extended region around the 
nucleus are clearly visible in the \ha map (Figure~\ref{f:ngc3735}).
\onlfig{16}{
\begin{figure*}
\centering
\includegraphics[scale=0.8]{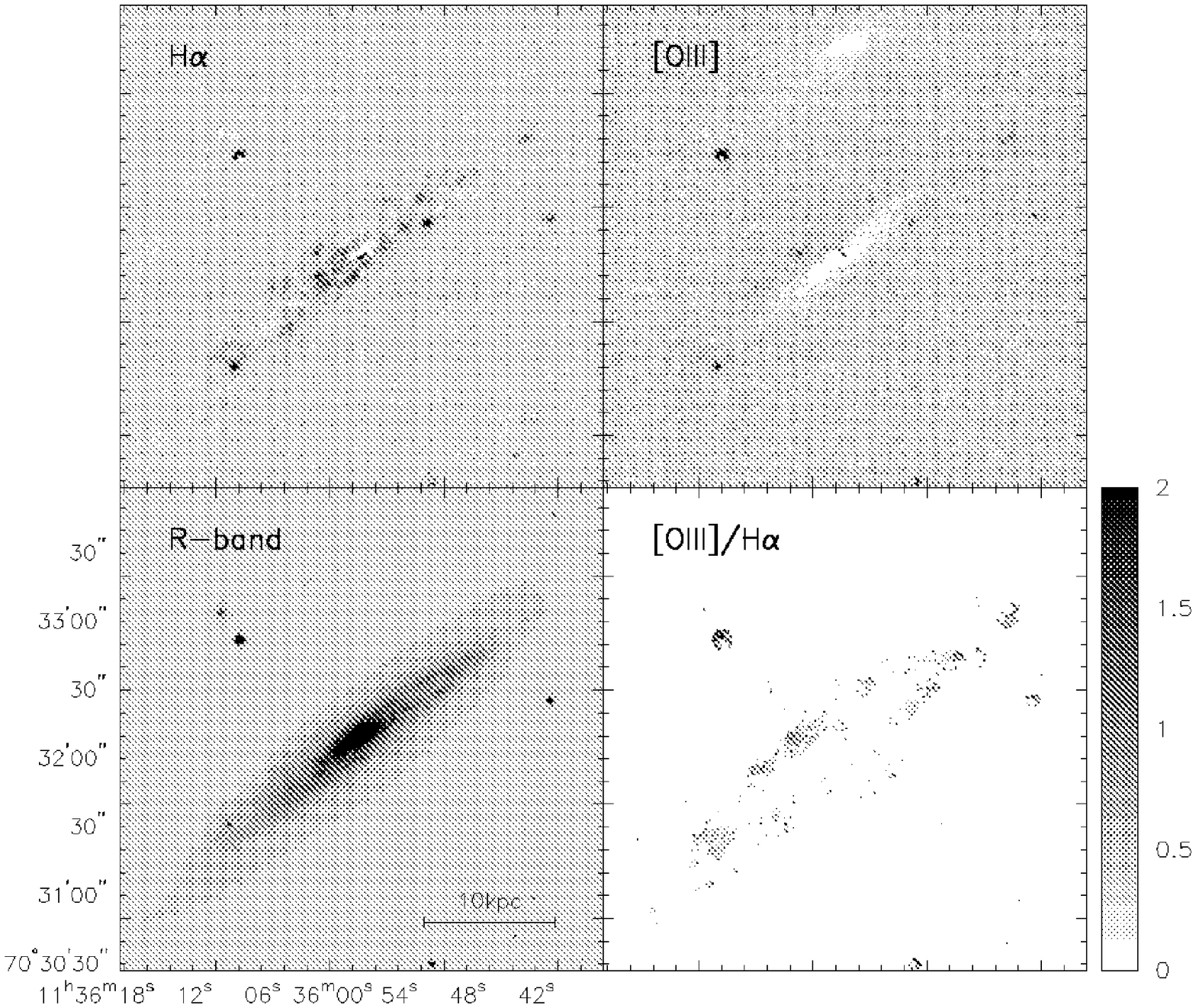}
\caption{Broadband, narrowband images and ionization map of NGC\,3735. \ha
(upper left), \oxy (upper right), $R$-band (lower left), and \oxy/\ha (lower
right). \label{f:ngc3735}}
\end{figure*}
}
 The \oxy 
map shows a few compact regions, most of which are off-planar and have
counterparts in the \ha 
map. As shown in high contrast \ha and \oxy 
maps (Figure~\ref{f:fourha}), possible extraplanar regions emitting in \oxy, 
but barely visible in \ha, are seen $38\arcsec$ (6.6~kpc) to the NE and 
$27\arcsec$ (4.7~kpc) to the SW of the nucleus. The high inclination of the 
galaxy makes these regions more likely to be extraplanar than in the galactic 
disk. The diffuse emission visible in the high-contrast \ha map could, 
combined with these extraplanar \oxy regions, be signs of minor-axis 
outflows.

\subsection{NGC\,4235}

The $R$-band image of NGC\,4235 (Figure~\ref{f:ngc4235})
\onlfig{17}{
\begin{figure*}
\centering
\includegraphics[scale=0.8]{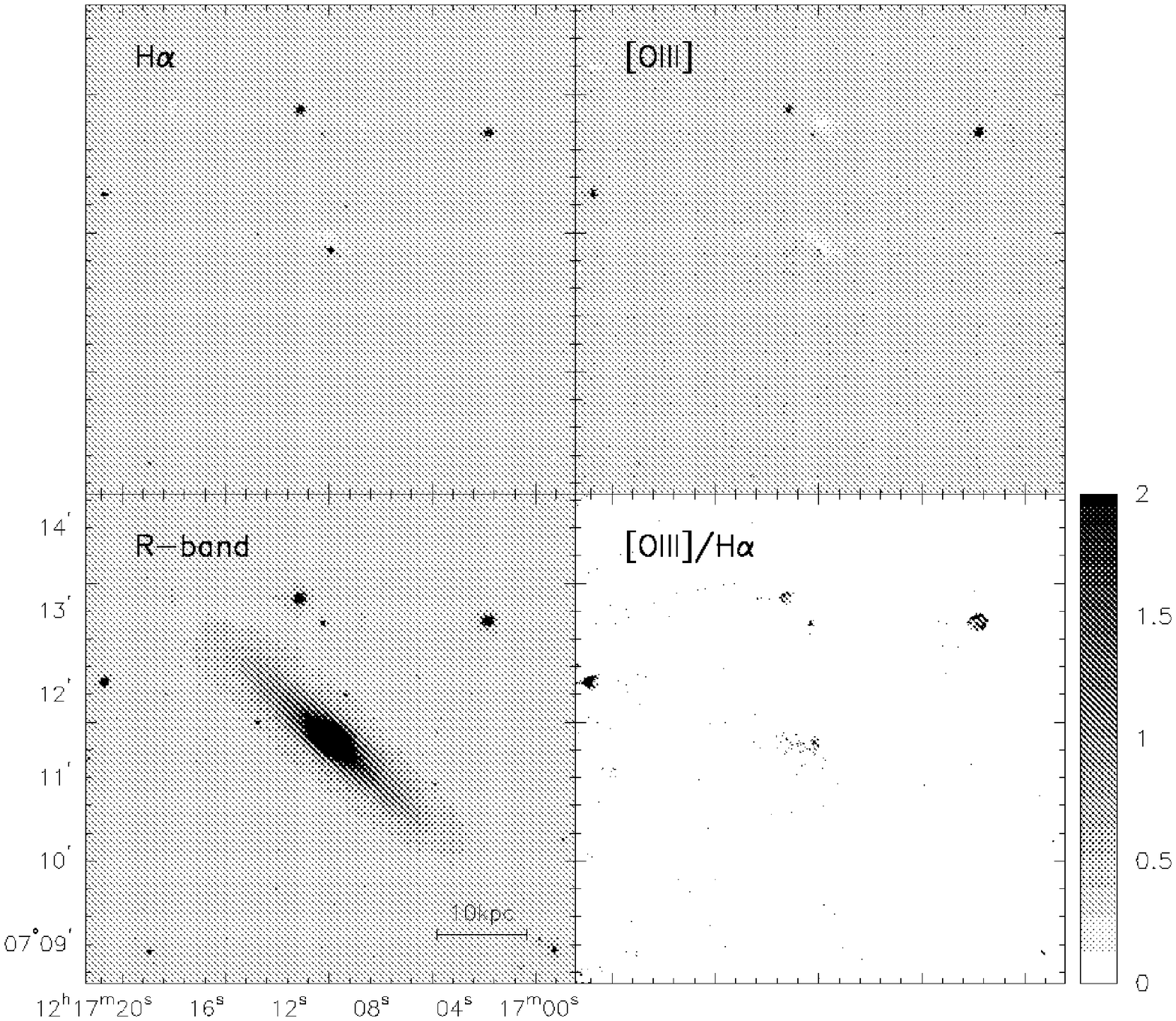}
\caption{Broadband, narrowband images and ionization map of NGC\,4235. \ha
(upper left), \oxy (upper right), $R$-band (lower left), and \oxy/\ha (lower
right). \label{f:ngc4235}}
\end{figure*}
}
 reveals a bright 
nuclear region, with a faint dust lane visible on either side of the nucleus. 
All of the \ha and \oxy emission originates from the nucleus. This is in 
agreement with previous observations \citep{pog89,col96,ros03b}. Absorption 
due to the dust lane is visible to the NW of the nucleus in both narrowband 
images. The absence of \hbox{H\,{\sc ii}} regions in the disk and the absence 
of extraplanar emission in both maps is in agreement with \cite{pog89}, 
although deeper observations by \citet{ros03b} reveal faint extended 
emission in \ha.

\subsection{NGC\,4388}

The $R$-band image of NGC\,4388 (Figure~\ref{f:ngc4388})
\onlfig{18}{
\begin{figure*}
\centering
\includegraphics[scale=0.8]{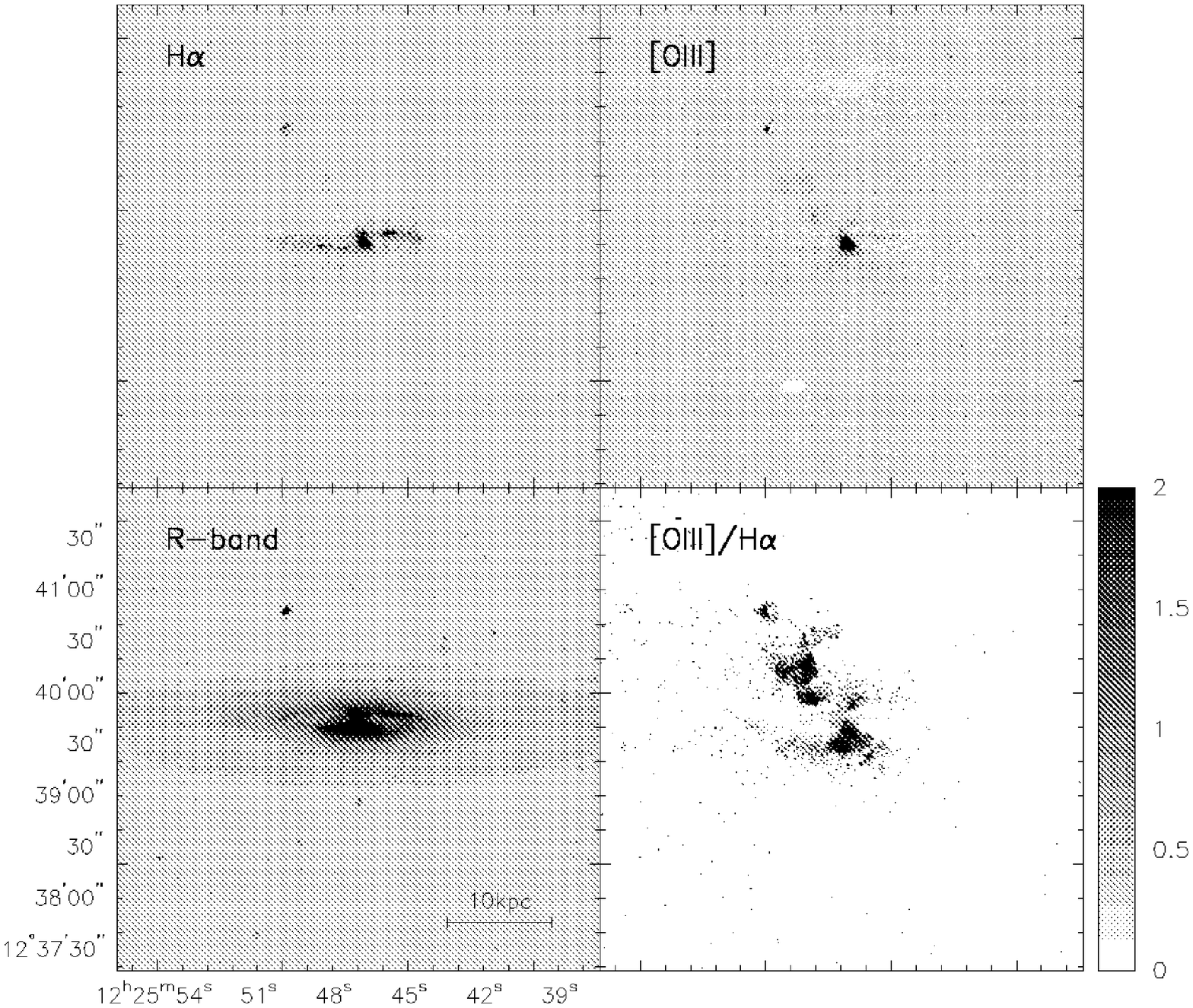}
\caption{Broadband, narrowband images and ionization map of NGC\,4388. \ha
(upper left), \oxy (upper right), $R$-band (lower left), and \oxy/\ha (lower
right). \label{f:ngc4388}}
\end{figure*}
}
 show a complicated
morphology, where the 
spiral arms are prominently featured in this highly inclined galaxy. Also 
present are several circumnuclear dusty regions. The narrowband images, 
however, show a much more intriguing morphology. NGC\,4388 is one of the best 
studied cases of minor axis outflows 
\citep[e.g.,][]{pog88b,pog89,vei99a,vei99b,yos02,vei03}. Our \ha map shows
evidence for spiral arms with a number of 
\hbox{H\,{\sc ii}} regions, as well as the nucleus whose morphology is 
reminiscent of a bar joining the base of the two spiral arms. Also visible in 
the high-contrast image shown in Figures~\ref{f:fourha} 
and~\ref{f:ngc4388zoom} (zoom-in) are the outflows extending $57\arcsec$ 
(4.6~kpc) to the N and $31\arcsec$ (2.5~kpc) to the S of the midplane of the 
galaxy. These are only small sections of the large-scale outflows which 
extend out to distances of 35~kpc from the nucleus \citep{yos02}. The \oxy 
map shows strong emission from the nucleus, which extends in the southern 
direction. Also visible are outflows to the N and S, corresponding to 
those seen in \ha, and diffuse emission from the western spiral arm. By and 
large, the nuclear region looks more complex in \oxy than it does in \ha.
\onlfig{19}{
\begin{figure*}
\centering
\includegraphics[scale=0.8]{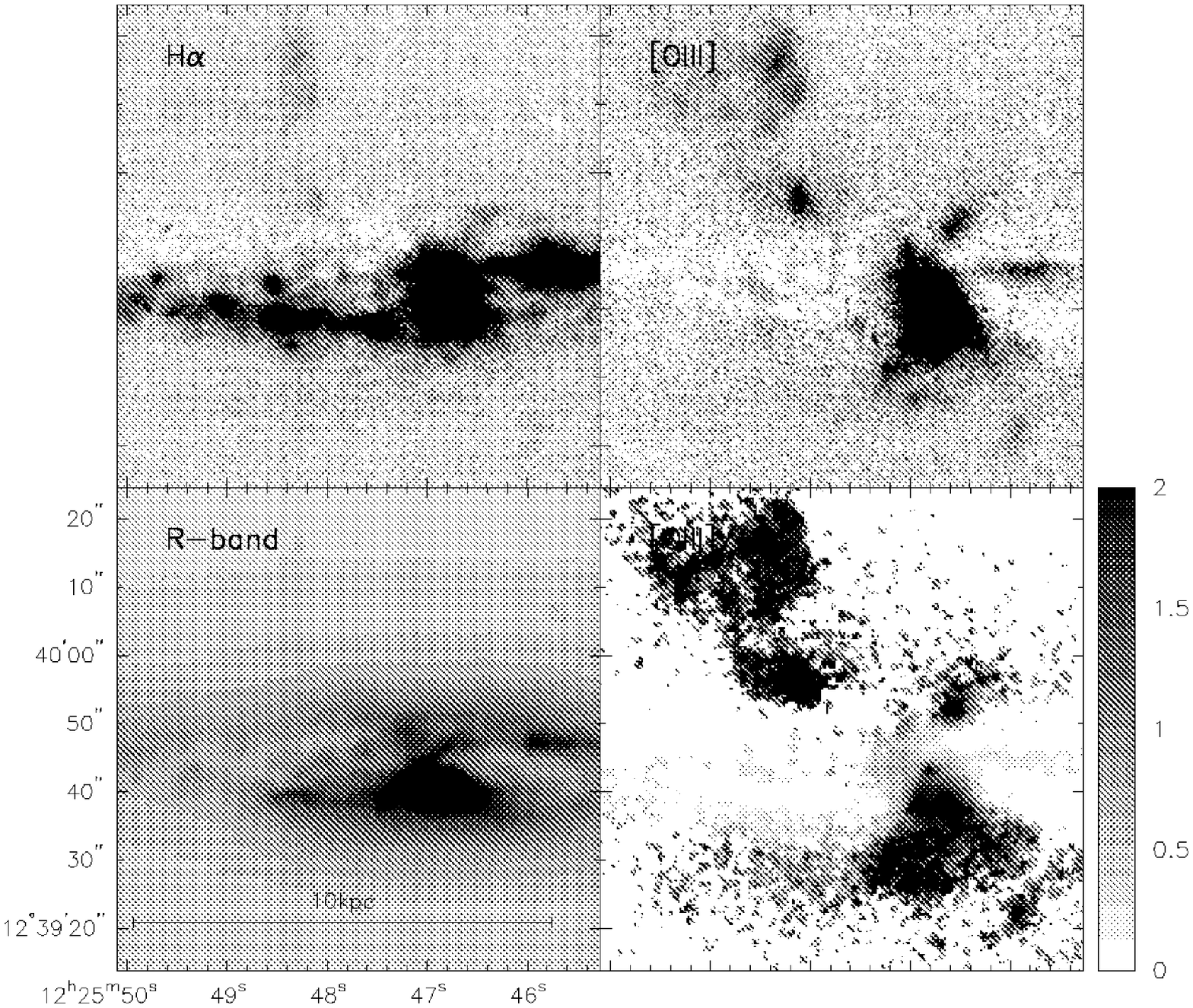}
\caption{Zoom-in version of NGC\,4388 in different contrast as
compared to Figure~\ref{f:ngc4388}. \label{f:ngc4388zoom}}
\end{figure*}
}

\subsection{NGC\,4565}

The $R$-band image of this highly-inclined Seyfert galaxy 
(Figure~\ref{f:ngc4565})
\onlfig{20}{
\begin{figure*}
\centering
\includegraphics[scale=0.8]{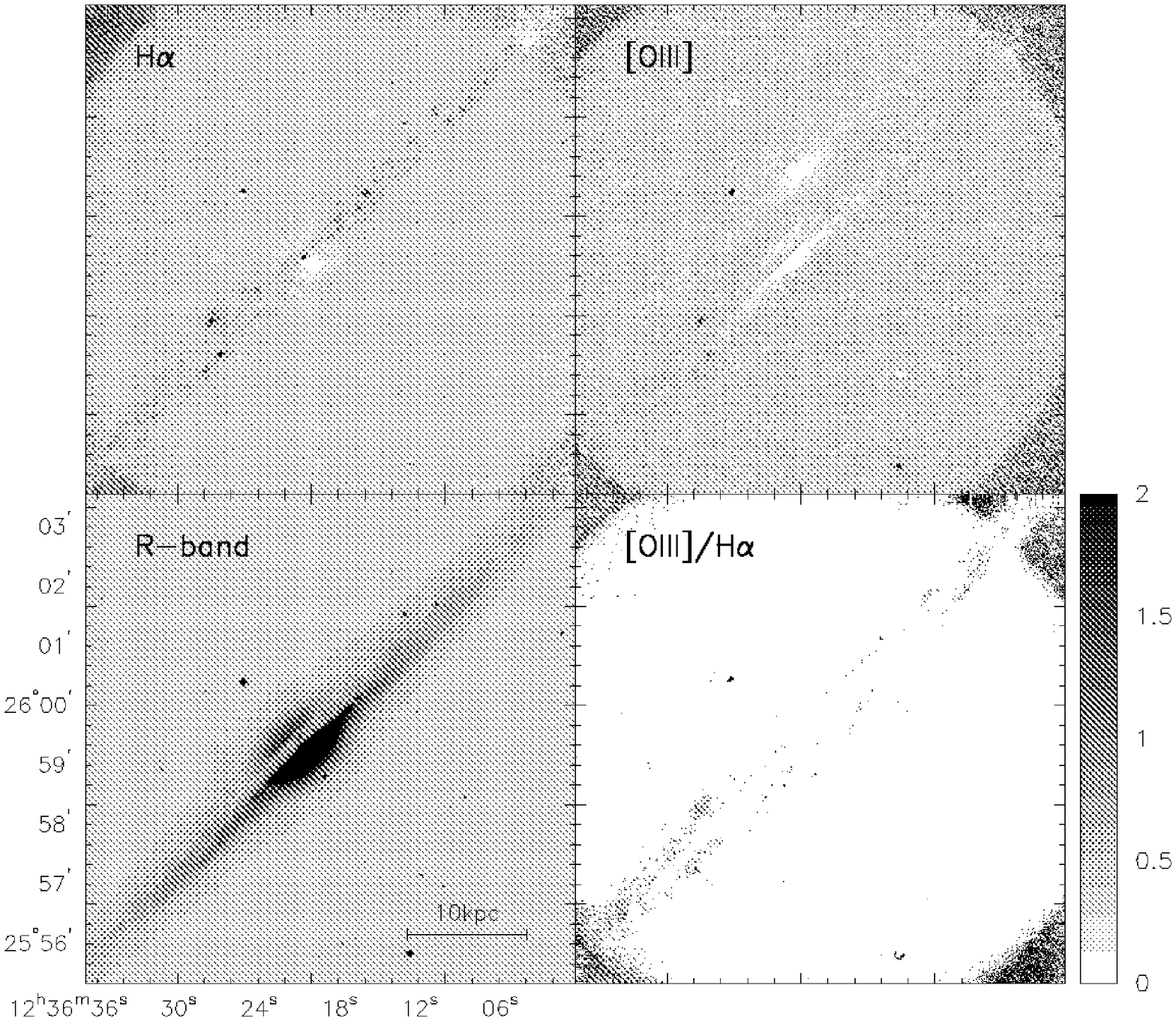}
\caption{Broadband, narrowband images and ionization map of NGC\,4565. \ha
(upper left), \oxy (upper right), $R$-band (lower left), and \oxy/\ha (lower
right). \label{f:ngc4565}}
\end{figure*}
\clearpage
}
 reveals a very prominent dust lane. Single 
chimney-like dust structures protrude from the northern plane perpendicular 
to the disk into the disk-halo interface. The \ha map of NGC\,4565 shows 
clumpy emission line regions and diffuse ionized gas scattered through the 
disk. No extraplanar \ha emission is observed, in agreement with previous 
observations performed by \citet{ran92}. However, there is possible evidence 
for some faint filaments. The brightest regions to the SE of the 
nucleus, as well as a few other emission regions in \ha have counterparts in 
the \oxy image. The disk region that displays the strongest \oxy emission, 
is shown in Figure~\ref{f:ngc4565z}. The brightest \oxy source is an 
off-planar \hbox{H\,{\sc ii}} region, located to the N of the disk. The
bulge is 
clearly not perfectly subtracted in the overall \oxy image, and this leads to 
some spuriously low values in the ionization map.
\onlfig{21}{
\begin{figure*}
\centering
\includegraphics[scale=0.8]{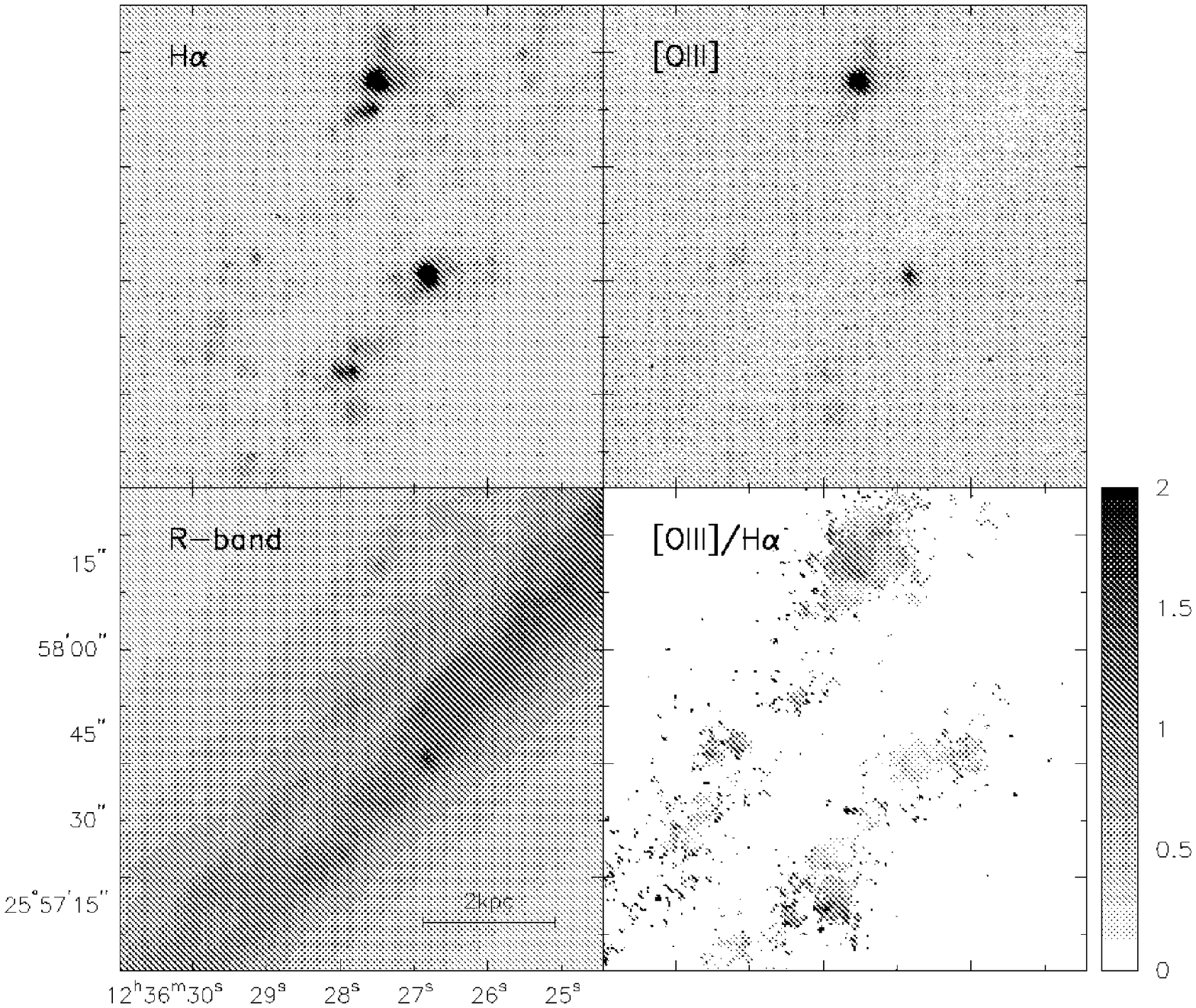}
\caption{A zoom-in of Figure~\ref{f:ngc4565} on the southern disk region
with prominent \hbox{H\,{\sc ii}} regions.\label{f:ngc4565z}}
\end{figure*}
}

\subsection{NGC\,5866}

The $R$-band image of NGC\,5866 (Figure~\ref{f:ngc5866})
\onlfig{22}{
\begin{figure*}
\centering
\includegraphics[scale=0.8]{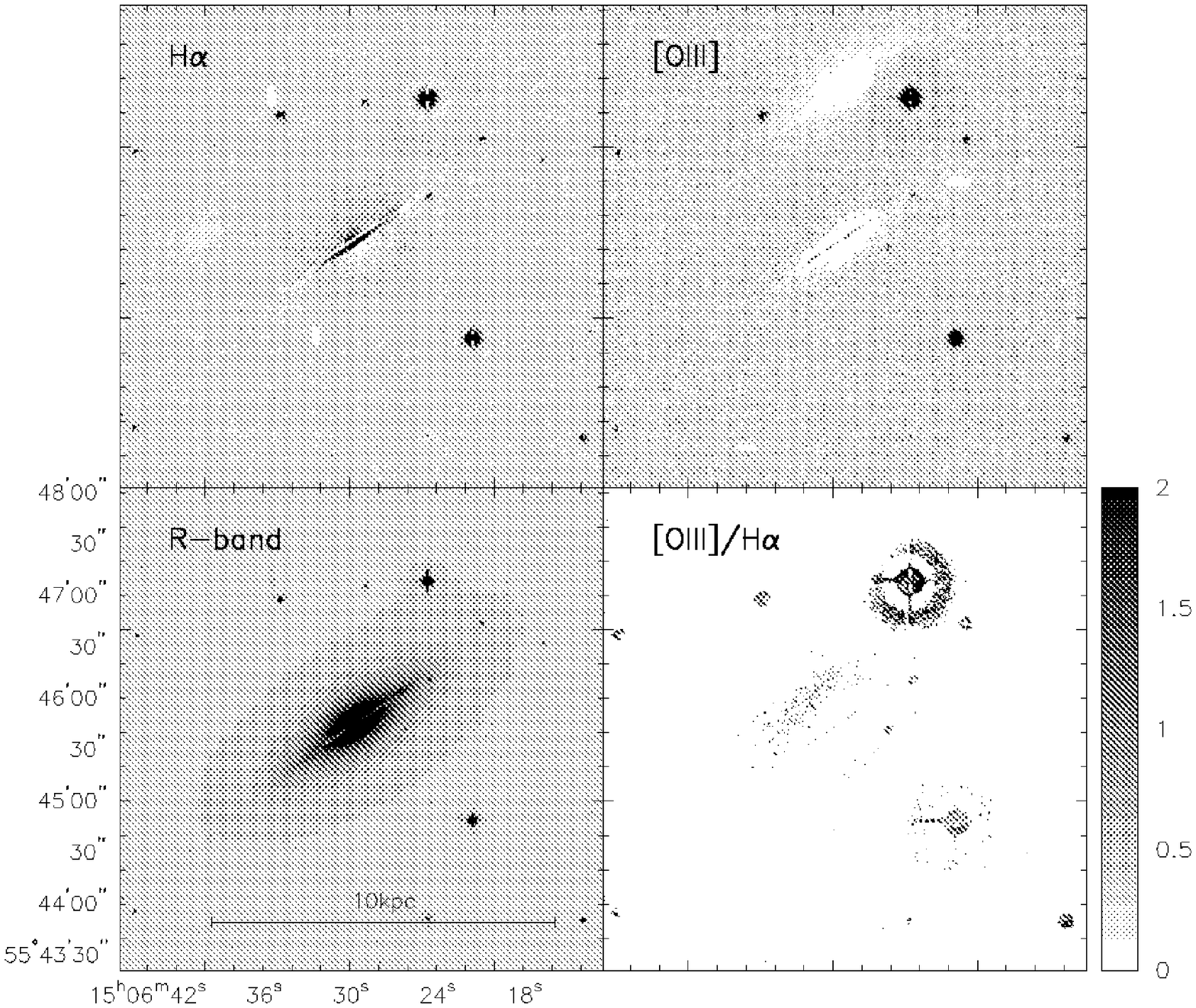}
\caption{Broadband, narrowband images and ionization map of NGC\,5866. \ha
(upper left), \oxy (upper right), $R$-band (lower left), and \oxy/\ha (lower
right). \label{f:ngc5866}}
\end{figure*}
}
 shows a prominent
bulge and a dust lane. Unlike other 
galaxies in our sample, no clumpy \hbox{H\,{\sc ii}} regions are seen in \ha.
The diffuse \ha emission is partially obscured by the prominent dust lane and
is 
asymmetric with respect to the plane of the disk. The \oxy emission is also 
partially obscured by the dust lane, and does not display any clumpy regions. 
It is not 
clear whether the diffuse emission seen on either side of the dust lane in 
\ha is a real feature and whether this contains evidence for any outflows. 
More likely its presence is artificial due to difficulties associated with 
the continuum subtraction.
The continuum subtraction proved very delicate for this galaxy and therefore 
we do not present any flux or ionization ratio measurements.

\subsection{IC\,1368}

The $R$-band image of IC\,1368 (Figure~\ref{f:ic1368}) shows an X-shaped region
of possibly clustered \hbox{H\,{\sc ii}} regions.
In agreement with \citet{col96}, we find that the \ha image 
shows a nucleus extended in the minor axis direction. 
Furthermore, we find evidence for a cone-like structure at the NW tip of the 
nucleus. In addition to faint diffuse emission from the disk, the \ha 
emission map shows possible evidence for two spiral arms containing a few 
bright \hbox{H\,{\sc ii}} regions. The \oxy image shows a bright nucleus 
surrounded by two strong absorption features, as well as diffuse disk 
emission.
\onlfig{23}{
\begin{figure*}
\centering
\includegraphics[scale=0.8]{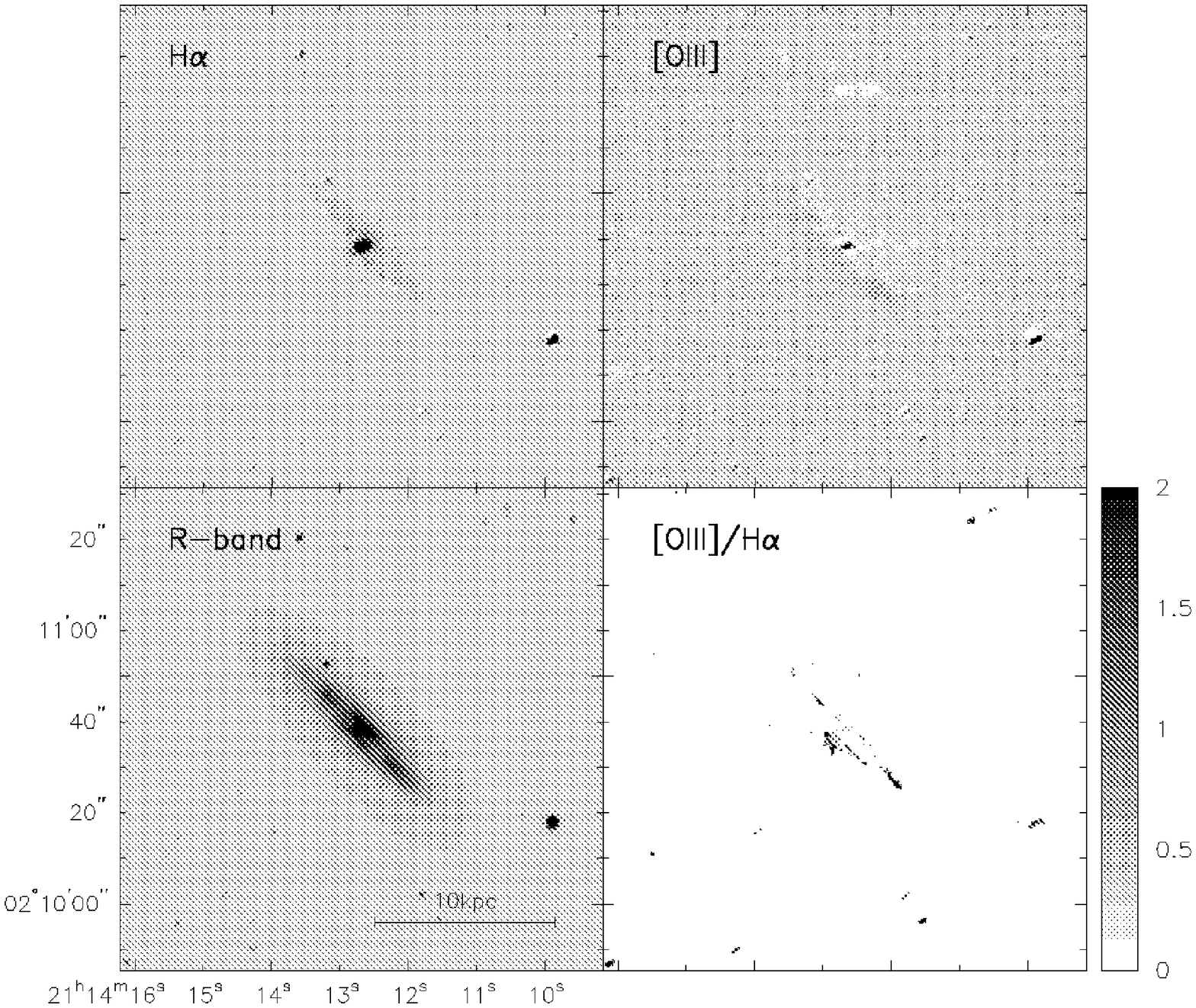}
\caption{Broadband, narrowband images and ionization map of IC\,1368. \ha
(upper left), \oxy (upper right), $R$-band (lower left), and \oxy/\ha (lower
right). \label{f:ic1368}}
\end{figure*}
}

\subsection{UGC\,12282}

The $R$-band image of UGC\,12282 (Figure~\ref{f:ugc12282}) shows some dusty
regions in between the modest 
\hbox{H\,{\sc ii}} regions outlining the spiral arms, which are partially 
visible in this inclined spiral galaxy. The \ha map shows
emission from \hbox{H\,{\sc ii}} regions 
scattered in the disk as well as from the nucleus. A ring of enhanced \ha 
emission with radius $23\arcsec$ (7.6~kpc) is also visible. The \oxy map does 
not show much emission outside the nucleus.
\onlfig{24}{
\begin{figure*}
\centering
\includegraphics[scale=0.8]{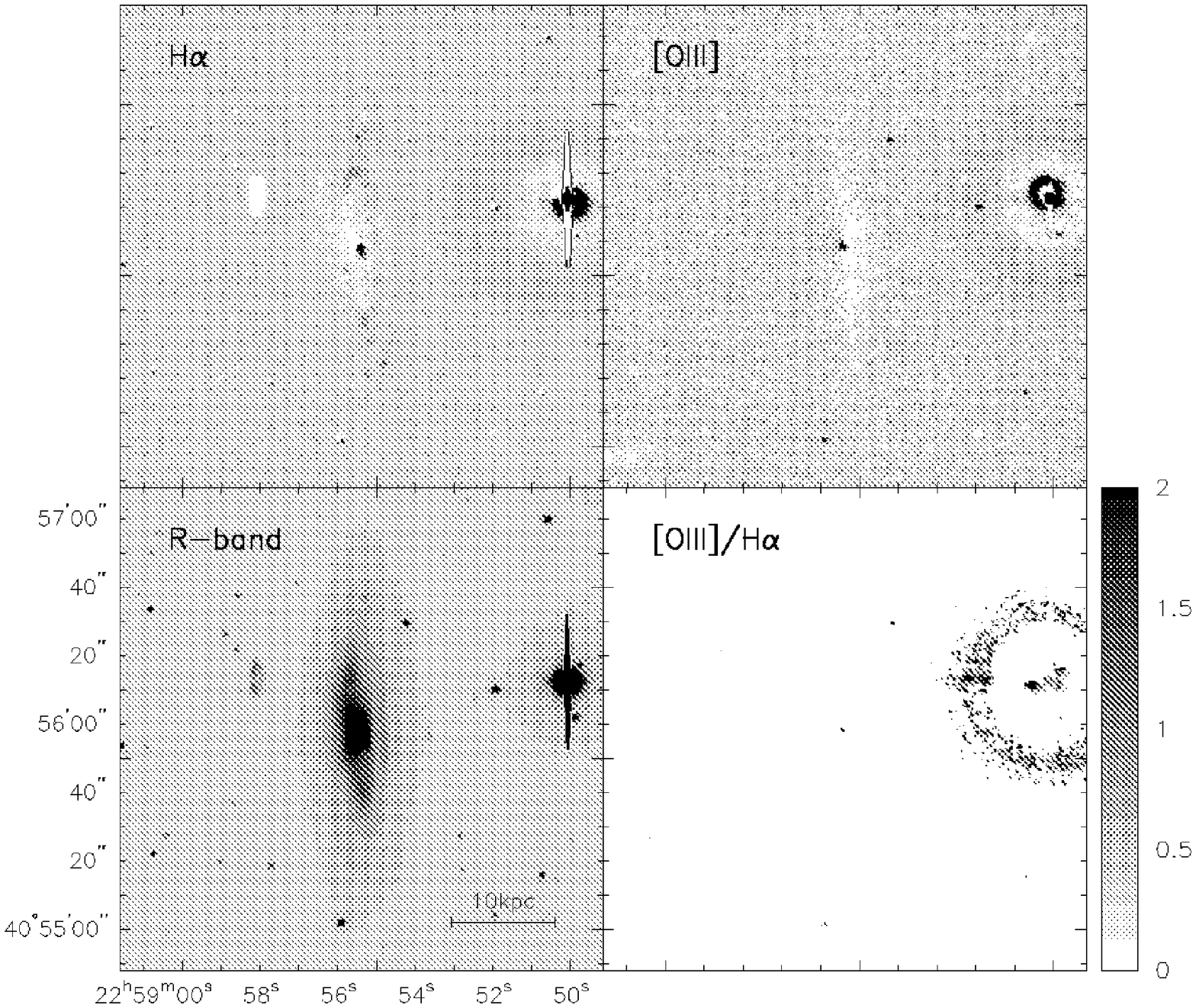}
\caption{Broadband, narrowband images and ionization map of UGC\,12282. \ha
(upper left), \oxy (upper right), $R$-band (lower left), and \oxy/\ha (lower
right). The elongated white/black feature in the $R$-band/\ha image to the
E of the disk is a ghost of the bright star close to the right corner of
the displayed field of view (see Section~\ref{ss:artifacts}).
\label{f:ugc12282}}
\end{figure*}
}
	
\section{Discussion}
\label{s:discuss}

\subsection{The ionization maps}
\label{ss:imapdiscuss}

One of the important diagnostics for unraveling the excitation mechanisms 
in galaxies is the use of emission lines of various excitation levels. In 
one-dimensional space (i.e. derived from longslit spectra) various 
measured line ratios (e.g., [\hbox{N\,{\sc ii}}]/\ha or \oxy/H$\beta$) 
can be used to discriminate between AGN, LINER, or \hbox{H\,{\sc ii}} 
emission line regions, making use of well known diagnostic diagrams 
\citep[e.g.,][]{vei87}. The advantage of emission line imaging is that it 
provides two-dimensional coverage. Therefore, we get an excitation 
map as a function of the position within the galaxy (i.e. for individual 
regions of interest) and for each covered pixel of the entire galaxy in 
general. The emission lines used in our study are some of the most prominent 
ones seen in Seyfert galaxies in the optical regime, \ha and \oxy. We 
created the \oxy/\ha ionization maps, shown in the lower right hand panels 
of Figures~\ref{f:mrk993} to \ref{f:ugc12282}. These maps might have been 
very useful in the characterization of any newly detected outflows. However, 
we did not find any new outflows but merely confirm the already well-studied 
outflows in NGC\,3079 and NGC\,4388. Therefore, our \oxy/\ha maps did not 
provide much new insight in the end, except in the cases which had previously 
only be studied in \ha. Nonetheless, we briefly discuss them here for 
completeness.  

The majority of our ionization maps are restricted to the nuclear and 
circumnuclear region, as our sensitivity did not reveal much \oxy emission 
within the disk. However, a few galaxies did reveal off-planar emission and 
disk emission line regions. In most cases, though we find that it is related 
to \hbox{H\,{\sc ii}} regions and not to diffuse emission originating from 
outflows. The sole exceptions are the well studied spirals NGC\,3079 and 
NGC\,4388. In the case of NGC\,3079 the nuclear superbubble is prominently 
visible in the \oxy/\ha ionization map, which has values of about 0.2 to 0.4. 
Values of the ratio are higher in the disk, where \oxy emission is quite 
strong locally, ranging from about 0.3 to 1.0, with peaks reaching 1.4 or 
more in some places. By and large the northern part of the disk has the 
strongest regions (high values of \oxy/\ha $\sim$ 0.7), whereas the southern 
part of the disk regions has considerably lower values (\oxy/\ha $\sim$ 0.3). 

In the case of NGC\,4388 the ionization map shows a `cone' of ionization to 
the S, in agreement with the measurements of \citet{pog88b}. Spectra and 
diagnostic line ratios presented by \citet{pog88b} showed that the nucleus 
and the circumnuclear regions can be described by photoionization of the 
central power-law continuum. In the ionization cone we find \oxy/\ha $\sim$ 
1.9 towards the center of the galaxy. It decreases to 1.4 and increases again 
to 2.4 further out from the center. The diffuse emission, visible in the two 
regions at greater extraplanar distances to the N, has much larger ionization 
values, roughly 2.6$\pm$0.7 in the upper region, and 4.3$\pm$0.9 in the lower 
region, both indicative of shock-ionization. 

In the case of Ark\,79, the \oxy/\ha ratio has a value of about 0.5 for the 
eastern main nucleus, and about 1.2 for the offset secondary nucleus. The 
nuclear region of Mrk\,1040 has a value of 0.2 to 0.3, with the most 
prominent region in the disk of this galaxy having similar ratios of 
0.3 to 0.4. This is lower than the \oxy/\ha ratio of the companion galaxy, 
which is 0.6.

The emission-line regions seen in the disk of NGC\,3735 all have values 
below 0.7. We note that the \oxy/\ha ratio in the possible extra-planar 
\hbox{H\,{\sc ii}} regions cannot be measured due to a too low 
signal-to-noise ratio. The \oxy/\ha ratio in the disk of NGC\,4565 does not 
exceed 1.0 in regions where \oxy emission is detected. The majority of the 
detected regions have \oxy/\ha $\leq$ 0.7.

The nuclear region of Mrk\,577 has \oxy/\ha values in the range of 0.8 to 
1.0. In Mrk\,993, the value of the ratio varies between 0.2 and 0.9 in the 
nuclear region, with the innermost region having values closer to 0.6. 
In UGC\,1479, the \oxy/\ha ratio varies between 1.0 and 1.4 in the nucleus. 
Finally, in UGC\,2936 the values of the ratio are close to 0.5 in 
the nuclear region. However, these values should be treated with caution, 
since the bright nucleus of some of the galaxies is the most sensitive 
region for the continuum subtraction, and the \oxy or \ha emission seen in 
the nuclei may be an artifact thereof. The nucleus of NGC\,4235 has 
\oxy/\ha $\sim$ 0.5, and this is the only region detected in the ratio map.

For NGC\,5866 and UGC\,12282 the nuclear regions proved unsuitable for the 
ionization map to permit any meaningful measurement of the ratio, and in 
IC\,1368 poor S/N also prevented any confident measurement.

\subsection{Nuclear versus minor axis disk outflows}
\label{ss:outflows}

\begin{figure*}
\centering
\includegraphics[angle=270,scale=0.7]{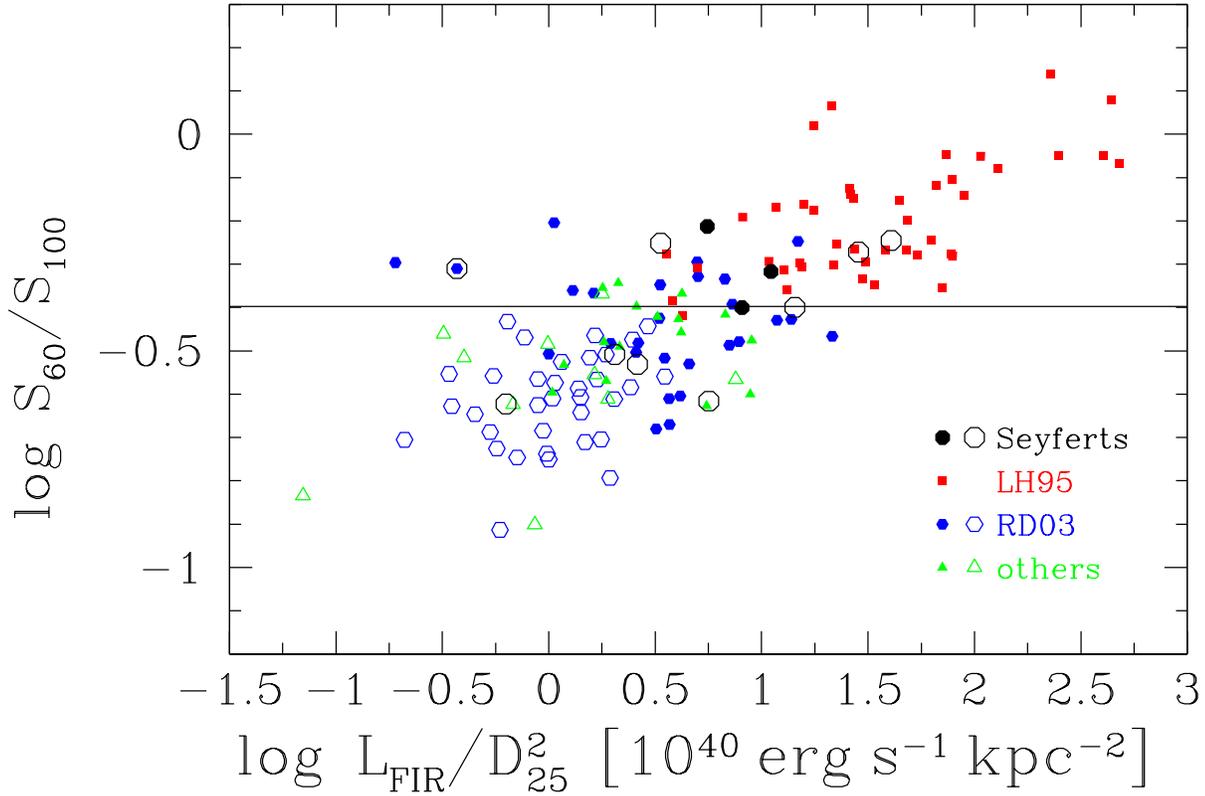}
\caption{Diagnostic diagram for extraplanar diffuse ionized gas, originating
from disk outflows into the halo of the galaxy, including the Seyfert
galaxies studied here (adapted and expanded from \citet{ros03a}). It shows
the star formation rate per unit area as a function of the IRAS flux ratio
($\log S_{60} / S_{100}$), which is a proxy for the warm dust temperature.
Filled symbols indicate galaxies with detected extraplanar
diffuse ionized gas, open symbols denote those without. For two of our
studied Seyfert galaxies (Mrk\,577 and Ark\,79) there were no FIR
measurements available. Therefore, no data points are plotted for these two
objects. The square symbols denote the starburst galaxies studied by
\citet{leh95}, the small hexagons denote non-starburst galaxies studied by
\citet{ros03a} and the triangles indicate data points for non-starburst
galaxies studied by others (see \citet{ros03a} for details). \label{f:ddd}}
\end{figure*}

The majority of the galaxies studied in this sample show strong
nuclear emission, as expected for Seyferts. All but two galaxies
(NGC\,4565 and NGC\,5866) show relatively strong nuclear emission in
\ha, and most in \oxy too. The two reported exceptions both have a
very prominent dust layer, which causes considerable extinction. Four
galaxies (Ark\,79, NGC\,3079, NGC\,4388 and IC\,1368) have complex
circumnuclear emission line regions. In contrast, UGC\,1479,
Mrk\,1040, NGC\,3079, NGC\,3735, NGC\,4388, NGC\,4565 and IC\,1368
show off-nuclear (extraplanar) emission line regions. However, not all 
(e.g.,\,UGC\,1479 and Mrk\,1040) show diffuse extraplanar emission
A comparison of the relation between star formation rate per unit area with 
dust temperature is shown in Figure~\ref{f:ddd}
(see \citet{ros03a} for
more general details). The Seyfert galaxies behave somewhat
differently in terms of the location within this diagram than the non-Seyfert 
galaxies. The majority of the Seyferts (nine versus three) for which FIR data 
are available, show strong nuclear emission, but have no extended diffuse 
ionized gas (DIG). This is true even at high FIR luminosity per unit area and 
high dust temperature. By contrast, in non-Seyfert galaxies this almost
uniquely indicates the presence of extraplanar DIG, expelled from the
active SF regions in the disk into the halo of these galaxies
\citep[e.g.,][]{leh95,ros03a,mil03}. This result is probably not an
artifact of differences in sensitivity for detection of DIG between
the present study and that of \citet{ros03a}. Both studies used
telescopes of similar size with similar observational parameters. Most
Seyfert galaxies studied here show no evidence for extended emission
line regions in both \ha and \oxy. The activity is solely restricted
to the nuclear and circumnuclear environment.  Only NGC\,3079,
NGC\,3735 and NGC\,4388 show extended DIG emission. NGC\,3079 is also
known to have a central starburst \citep{cec02} in addition to its
Seyfert nucleus, and that may explain why extended emission is
detected. NGC\,4388 is a galaxy experiencing ram pressure stripping in
the Virgo cluster environment \citep{ken04}, which may be a primary
contributor to its very extended emission line regions especially far
from the nucleus. So overall it appears that star formation is the
primary mechanism producing extended DIG around the spiral galaxies
studied by the present study and that of \citet{ros03a}. In general,
AGN activity undoubtedly plays some role in driving minor-axis
outflows. However, this probably requires higher luminosities than are
encountered in our small distance-limited sample.

For Ark\,79, which was classified as a type 2 Seyfert galaxy \citep{ost77}, 
we find evidence for a secondary nucleus with a separation of 
$\approx 2\farcs5$, which is clearly visible in the \oxy image, but only 
faintly in the \ha image. The two detected components are aligned roughly 
along the major axis of the disk. As it is barely detected in \ha, the 
secondary component was undetected in previous studies 
\citep[e.g.,][]{col96}. The projected distance corresponds roughly to 
850\,pc. This is the only case in our sample which shows a two-component 
structure of the nuclear region. Apart from this we find at least two more 
cases with extended circumnuclear regions. NGC\,4388 shows a relatively 
complex circumnuclear region, specifically in \oxy, in agreement with 
previous studies \citep[e.g.,][]{vei99a,yos02}. An extended cone-like nuclear 
outflow region is detected in IC\,1368. This is reminiscent of a downsized 
version of a prominent starburst cone. 
    
\section{Summary}
\label{s:sum}

We have observed a distance limited sample of 14 nearby edge-on
Seyfert galaxies with narrowband imaging in \ha and \oxy to study the
role of the active nucleus and SF regions within the disk for the IGM
enrichment. Because of the distance-limited nature of the sample, it
is restricted to relatively low-luminosity Seyfert galaxies. The
median 3-$\sigma$ sensitivities for detection of high-latitude
extended emission in the sample galaxies are
$3.6\times10^{-17}$\,ergs\,s$^{-1}$\,cm$^{-2}$\,arcsec$^{-2}$ for the
\ha images and
$6.9\times10^{-17}$\,ergs\,s$^{-1}$\,cm$^{-2}$\,arcsec$^{-2}$ for the
\oxy images. The studied galaxies show a variety of emission line
features. The majority, however, reveal only emission in the nuclear
and circumnuclear environment at the sensitivity of our
observations. Extraplanar emission from strong SF regions in the disk
is only observed in three cases (NGC\,3079, NGC\,3735 and NGC\,4388),
with filaments reaching distances of up to 13.5\,kpc above/below the
disk in NGC\,3079. Single very faint filaments also seem to originate
from strong SF disk regions in NGC\,4565, although only to much lower
galactic latitudes. In Ark\,79 we detect a secondary nuclear
component, most prominently seen in the \oxy image. The two components
are separated by 850\,pc.
 
Whereas the \ha emission often shows a variety of discrete features
(filaments, emission line cones, \hbox{H\,\sc{ii}} regions), the \oxy
emission distribution is, with very few exceptions, mainly restricted
to the nucleus and in a few cases to the circumnuclear
environment. This is probably due to our limited sensitivity, combined
with the fact that extended gas, except in very exceptional
circumstances, is expected to be relatively low-excitation (based on
ionization parameter arguments considering low-density gas very far
from an AGN). In total we detected large scale minor axis outflows
only in three of the studied 14 galaxies. The corresponding detection
rate of 21\% is similar to the detection rate found by \citet{col96},
which is 27\%. The study of \citet{vei03} had higher sensitivity than
both of these studies. However, the results of this latter paper
cannot be used for statistical purposes because their sample was
pre-selected to contain objects with known ionization cones and/or
galactic winds on a roughly kiloparsec scale.

Overall, our results show that extraplanar emission of similar
brightness and extent as in the previously known cases of NGC\,3079
and NGC\,4388 is not common in Seyfert galaxies of otherwise similar
properties. In both these galaxies there are external effects, such as
a starburst wind or ram pressure stripping, that might effect the
extended emission. Comparison with the results of \citet{ros03a} shows
that for nearby edge-on spiral galaxies star formation may be a more
powerful mechanism for producing DIG than AGN activity. In general,
AGN activity undoubtedly plays some role in driving minor-axis
outflows. However, this probably requires higher AGN luminosities than
are enountered in our small distance-limited sample.  Future studies
with larger telescopes and better sensitivity will be helpful to shed
more light onto the role of minor axis disk outflows in edge-on
Seyfert galaxies.
 

\begin{acknowledgements} 
DJB and JR acknowledge financial support for this project by the DFG
through grants BO\,1642/2-1 and BO\,1642/3-1. TR wishes to thank the
Space Telescope Science Institute for repeated hospitality in hosting
him as both a summer student and a visitor. We thank the anonymous
referee for suggestions that helped improve our paper. This research
has made use of the NASA/IPAC Extragalactic Database (NED) which is
operated by the Jet Propulsion Laboratory, California Institute of
Technology, under contract with the National Aeronautics and Space
Administration.
\end{acknowledgements}



\begin{thebibliography}{}

\bibitem[Cayatte et al.(1994)]{cay94} Cayatte, V., Kotanyi, C., 
Balkowski, C., \& van Gorkom, J. H. 1994, \aj, 107, 1003

\bibitem[Cecil et al.(2001)]{cec01} Cecil, G., Bland-Hawthorn, J., Veilleux, 
S., \& Filippenko, A. V. 2001, \apj, 555, 338

\bibitem[Cecil et al.(2002)]{cec02} Cecil, G., Bland-Hawthorn, J., \& 
Veilleux, S. 2002, \apj, 576, 745

\bibitem[Colbert et al.(1996)]{col96} Colbert, E. J. M., Baum, S. A., 
Gallimore, J. F., et al. 1996, \apjs, 105, 75

\bibitem[Dahlem et al.(1997)]{dah97} Dahlem, M., Petr, M. G., Lehnert, 
M. D., Heckman, T. M., \& Ehle, M. 1997, \aap, 320, 731 

\bibitem[Heckman et al.(1990)]{hec90} Heckman, T. M., Armus, L., \& Miley, 
G. K. 1990, \apjs, 74, 833

\bibitem[Kenney \& Koopmann(1999)]{ken99} Kenney, J. D. P., \& Koopmann, 
R. A. 1999, \aj, 117, 181

\bibitem[Kenney et al.(2004)]{ken04} Kenney, J. D. P., van Gorkom, J. H., \& 
Vollmer, B. 2004, \aj, 127, 3361

\bibitem[Landolt(1992)]{lan92} Landolt, A. U. 1992, \aj, 104, 340 

\bibitem[Lehnert \& Heckman(1995)]{leh95} Lehnert, M. D., \& Heckman, T. M. 
1995, \apjs, 97, 89

\bibitem[Lehnert \& Heckman(1996a)]{leh96a} Lehnert, M. D., \& Heckman, 
T. M. 1996a, \apj, 462, 651

\bibitem[Lehnert \& Heckman(1996b)]{leh96b} Lehnert, M. D., \& Heckman, 
T. M.\ 1996b, \apj, 472, 546 

\bibitem[Miller \& Veilleux(2003)]{mil03} Miller, S. T., \& Veilleux, S. 
2003, \apjs, 148, 383

\bibitem[Osterbrock \& Phillips(1977)]{ost77} Osterbrock, D. E., \& 
Phillips, M. M. 1977, \pasp, 89, 251

\bibitem[Pogge(1988a)]{pog88a} Pogge, R. W. 1988a, \apj, 328, 519

\bibitem[Pogge(1988b)]{pog88b} Pogge, R. W. 1988b, \apj, 332, 702

\bibitem[Pogge(1989)]{pog89} Pogge, R. W. 1989, \apj, 345, 730

\bibitem[Rand(1996)]{ran96} Rand, R. J. 1996, \apj, 462, 712

\bibitem[Rand et al.(1992)]{ran92} Rand, R. J., Kulkarni, S. R., \& 
Hester, J. J. 1992, \apj, 396, 97

\bibitem[Reif et al.(1999)]{rei99} Reif, K., Bagschik K., de Boer K. S., 
et al. 1999, SPIE, 3649, 109

\bibitem[Reynolds (1992)]{rey92} Reynolds, R. J. 1992, ApJ, 392, L35

\bibitem[Rossa \& Dettmar(2000)]{ros00} Rossa, J., \& Dettmar, R.-J. 2000, 
\aap, 359, 433

\bibitem[Rossa \& Dettmar(2003a)]{ros03a} Rossa, J., \& Dettmar, R.-J. 
2003a, \aap, 406, 493

\bibitem[Rossa \& Dettmar(2003b)]{ros03b} Rossa, J., \& Dettmar, R.-J. 
2003b, \aap, 406, 505

\bibitem[Strickland et al.(2004)]{str04} Strickland, D. K., Heckman, T. M., 
Colbert, E. J. M., Hoopes, C. G., \& Weaver, K. A. 2004, \apjs, 151, 193

\bibitem[Veilleux \& Osterbrock(1987)]{vei87} Veilleux, S., \& Osterbrock, 
D. E. 1987, \apjs, 63, 295

\bibitem[Veilleux \& Rupke(2002)]{vei02} Veilleux, S., \& Rupke, D. S. 
2002, \apj, 565, L63

\bibitem[Veilleux et al.(1994)]{vei94} Veilleux, S., Cecil, G., 
Bland-Hawthorn, J., Tully, R. B., Filippenko, A. V., \& Sargent, W. L. W. 
1994, \apj, 433, 48

\bibitem[Veilleux et al.(1995)]{vei95} Veilleux, S., Cecil, G., \& 
Bland-Hawthorn, J. 1995, \apj, 445, 152

\bibitem[Veilleux et al.(1999a)]{vei99a} Veilleux, S., Bland-Hawthorn, J., 
Cecil, G., Tully, R. B., \& Miller, S. T. 1999, \apj, 520, 111

\bibitem[Veilleux, Bland-Hawthorn, \& Cecil(1999b)]{vei99b} Veilleux, S.,
Bland-Hawthorn, J., \& Cecil, G. 1999, \aj, 118, 2108

\bibitem[Veilleux et al.(2003)]{vei03} Veilleux, S., Shopbell, P. L., Rupke, 
D. S., Bland-Hawthorn, J., \& Cecil, G. 2003, \aj, 126, 2185

\bibitem[Veilleux et al.(2005)]{vei05} Veilleux, S., Cecil, G., \& 
Bland-Hawthorn, J. 2005, \araa, 43, 769

\bibitem[V\'eron-Cetty \& V\'eron(2001)]{ver01} V\'eron-Cetty, M. P., \& 
V\'eron, P. 2001, \aap, 374, 92

\bibitem[Walker et al.(1996)]{wal96} Walker, I. R., Mihos, J. C., \& 
Hernquist, L. 1996, \apj, 460, 121

\bibitem[Yoshida et al.(2002)]{yos02} Yoshida, M., Yagi, M., Okamura, S., 
et al. 2002, \apj, 567, 118

\bibitem[Yoshida et al.(2004)]{yos04} Yoshida, M., Ohyama, Y., Iye, M., 
et al. 2004, \aj, 127, 90

\end{thebibliography}
\end{document}